\documentclass[preprint,12pt]{elsarticle}

\usepackage{amssymb}
\usepackage[margin=3cm]{geometry}
\usepackage{setspace}
\usepackage{amsthm}
\usepackage{bm}
\journal{arXiv}
\usepackage{color}
\begin{document}

\begin{frontmatter}
\biboptions{numbers,sort&compress}

\title{\textit{Ab initio} investigation of the crystallization mechanism of cadmium selenide}

\author[label1]{Linshuang Zhang}
\affiliation[label1]{organization={State Key Laboratory of Solidification Processing, International Center for Materials Discovery, School of Materials Science and Engineering, Northwestern Polytechnical University},
            city={Xi'an},
            postcode={710072}, 
            country={China}}
\author[label2]{Manyi Yang\corref{cor1}}
\ead{manyi.yang@iit.it}
\affiliation[label2]{organization={Italian Institute of Technology},
            addressline={Via E. Melen 83}, 
            city={Genoa},
            postcode={16152}, 
            country={Italy}}
\author[label1]{Shiwei Zhang}
\author[label1]{Haiyang Niu\corref{cor1}}
\ead{haiyang.niu@nwpu.edu.cn}
\cortext[cor1]{Corresponding author}

\begin{abstract}
Cadmium selenide (CdSe) is an inorganic semiconductor with unique optical and electronic properties that made it useful in various applications, including solar cells, light-emitting diodes, and biofluorescent tagging. In order to synthesize high-quality crystals and subsequently integrate them
into devices, it is crucial to understand the atomic scale crystallization mechanism of CdSe. Unfortunately, such studies are still absent in the literature. 
To overcome this limitation, we employed an enhanced sampling-accelerated active learning approach to construct a deep
neural potential with \textit{ab initio} accuracy for studying the crystallization of CdSe.
Our brute-force molecular dynamics simulations revealed that a spherical-like nucleus formed spontaneously and stochastically, resulting in a stacking disordered structure where the competition between hexagonal wurtzite and cubic zinc blende polymorphs is temperature-dependent. We found that pure hexagonal crystal can only be obtained approximately above 1430 K, which is 35 K below its melting temperature. Furthermore, we observed that the solidification dynamics of Cd and Se atoms were distinct due to their different diffusion coefficients. The solidification process was initiated by lower mobile Se atoms forming tetrahedral frameworks, followed by Cd atoms occupying these tetrahedral centers and settling down until the third-shell neighbor of Se atoms sited on their lattice positions. Therefore, the medium-range ordering of Se atoms governs the crystallization process of CdSe. Our findings indicate that understanding the complex dynamical process is the key to
comprehending the crystallization mechanism of compounds like CdSe, and can shed lights in the synthesis of high-quality crystals.
\end{abstract}

\end{frontmatter}

\section{Introduction}
Cadmium selenide (CdSe) is a binary chalcogenide compound that has garnered extensive attention due to its unique nonlinear optical and wave absorbing  properties~\cite{Ninomyia1995,Park1993,yuan2016,yuan2017}. When confined to the nanoscale, CdSe exhibits tunable quantum confinement effects, and its quantum dots have been utilized in a broad range of applications, including solar cells~\cite{Greenhem1996,schierhorn2009}, light-emitting diodes~\cite{Schlamp1997,Mattoussi}, and biofluorescent tagging~\cite{Marcel1998,Warren1998}. In practical applications, a high-quality CdSe crystal with uniform properties is desired. However, synthesizing CdSe crystals is challenging due to their high melting temperature and high vapour pressure~\cite{steininger1968}. For its bulk phase, supercooling not far from the melting temperature is usually adopted in the Bridgman method to obtain a pure single crystal~\cite{ni2018,kolesnikov2003}. While for its nanoscale form, CdSe mainly crystallizes into a stacking disordered structure in a solution at relatively low temperature~\cite{hughes_anisotropic_2013,qu_alternative_2001}. Therefore, understanding the crystallization process of CdSe is essential to control and improve the quality and subsequently integrate it into more complex systems and advanced devices.

At ambient pressure, CdSe crystallizes into two different phases, namely hexagonal wurtzite (WZ) and cubic zinc blende (ZB)~\cite{Yeh1992,huang2010}, which are arranged in the order of ABAB... and ABCABC... sequence along the \textit{z}-axis, respectively (see Fig. S1). Both structures feature a tetrahedral diamond lattice network, one of the most ubiquitous structural prototypes~\cite{pauling2013}. Understanding the crystallization process of the diamond lattice-based structure has attracted extensive attention in the literature from the crystallization of silicon~\cite{2018silicon} to the homogeneous nucleation of ice and silica~\cite{niusilica2018,niu2019,chen2023}. However, to the best of our knowledge, studies on the nucleation process of two-component based diamond lattice structures, such as CdSe, are still absent in the literature. It is difficult to imagine how these two types of elements, i.e., Cd and Se, interact with each other and crystallize into a diamond lattice.

Molecular dynamics (MD) is a powerful  methodology to simulate the crystallization process of CdSe but is fraught with great challenges from several perspectives. For one thing, as a binary chalcogenide semiconductor compound, CdSe has both ionic and covalent characteristics~\cite{guo2013,chanda2020}, and chemical bonding changes need an accurate description of the atomic interaction. In addition, the ground state energy of cubic ZB phase is only 0.9 meV/atom lower than that of the hexagonal WZ phase based on first-principles calculations, while at relatively high temperatures the hexagonal phase is more stable~\cite{deng_new_2005}. 
In addition, the [001] face of hexagonal phase is indistinguishable from the [111] face of cubic phase, and a stacking disordered structure consisting of staggered layers of these two phases is  formed easily. Thus, the subtle difference between the hexagonal and cubic structures needs to be described accurately. \textit{Ab initio} MD (AIMD) is, in principle, a good solution to overcome these challenges. However, the high computational cost makes it prohibitive to perform large-scale and long simulations, which are essential to study the nucleation and growth processes of CdSe. 

 A way of circumventing these limitations while retaining the \textit{ab initio} accuracy is to follow the strategy pioneered by Behler and Parrinello~\cite{Behler2007,behler2016}, in which a feed-forward deep neural network (DNN) was trained to represent the potential energy surface. By fitting data sets from the electronic structure calculations of a large number of selected configurations and optimizing parameters, a DNN potential that can provide a direct functional relationship between the atomic configuration and \textit{ab initio} energy and atomic forces at a much lower computational cost can be obtained. This strategy has become very popular, and subsequently, some new methods such as DeePMD-kit~\cite{wang2018} have been developed. Extensive processes have been achieved for many complex dynamical phenomena ~\cite{niugallium2020,yang2021,yang2022,zhangprl,n2o5}, such as estimating the complex phase diagram of gallium~\cite{niugallium2020}, studying the liquid-liquid phase transition of phosphorus~\cite{yang2021} and simulating the homogeneous nucleation process of ice~\cite{chen2023}, which were thought to be impossible to investigate with \textit{ab initio} accuracy before.

Following the same spirit, in this work, we have combined the advanced enhanced sampling method~\cite{laio2002} with the DeePMD approach~\cite{zhang2018,wang2018} and built an \textit{ab initio} accuracy DNN potential that is capable of describing the complex behaviours of the hexagonal WZ, cubic ZB and liquid phases of CdSe. With this DNN potential, we were able to systematically study the nucleation and growth processes of CdSe. 

\section{Method}
\subsection{Training neural network potential}
We constructed the DNN potential following the Deep Potential MD method recently developed by Zhang et al.~\cite{wang2018}. In this method, the potential energy $E$ of each atomic configuration with $N$ atoms was expressed as $E = \sum_i^N E_i$, where $E_i$ is the atomic energy of atom $i$ depending on its atomic type and the local environments within a smooth cutoff radius $R_c$. The parameters of the DNN potentials were optimized on a relatively large set of density functional theory (DFT) energies, atomic forces, and virials performed on reference configurations of the training set. We refer the reader to Ref~\cite{wang2018} for further details. 
  
\begin{figure}[!ht]
  \centering
  \includegraphics[width=0.95\textwidth]{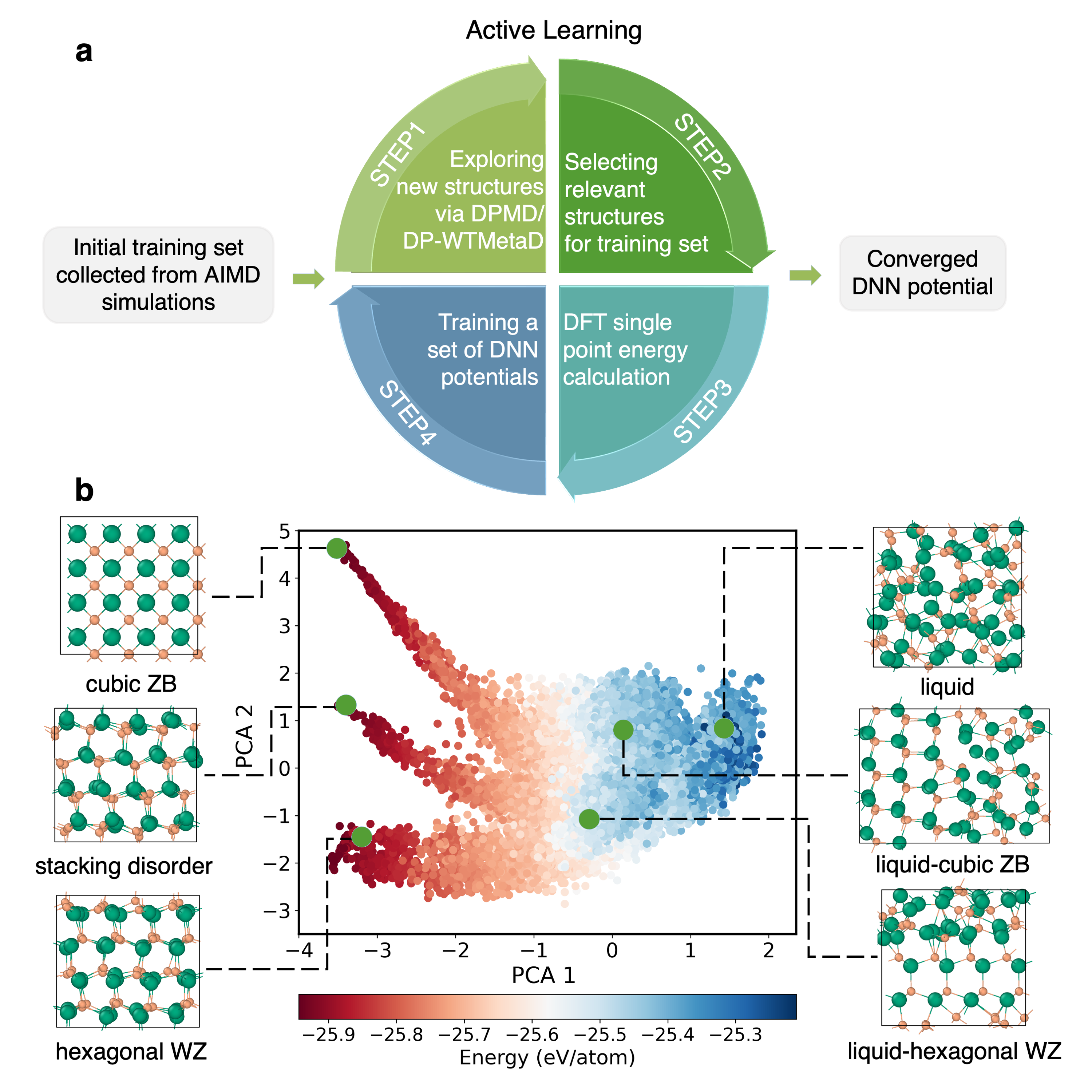}
  \caption{DNN potential training procedure and the distribution analysis of the training set. (a) Flowchart of the active 
  learning procedure to obtain the DNN potential.
  (b) The distribution of training set configurations in the two-dimension space mapped with the principal component analysis (PCA). The energy of each configuration is colored from red (-26.0 eV/atom) to blue (-25.2 eV/atom). Typical structures of pure liquid, hexagonal wurtzite, cubic zinc blende, stacking disordered, liquid-hexagonal wurtzite, and liquid-cubic zinc blende coexisting configurations are shown, respectively. In all snapshots of this work, the green and brown spheres refer to Se and Cd atoms whose sizes are proportional to the radii of the corresponding ions, respectively. 
  }
  \label{F:NN_train}
\end{figure}

The collection of the reference configurations relevant to the study of interest is a crucial step for the successful construction of the DNN potential. In studying the nucleation and growth of CdSe, this is made harder by the fact that we have to explore in the training set a vast configurational space associated with the reactive liquid-solid phase transition process, which includes all possible liquid, solids (hexagonal WZ, cubic ZB, and their stacking disordered phase), and many rare but critical liquid‒solid coexisting states of CdSe, as shown in Fig.~\ref{F:NN_train}b. To address this issue, we adopted the active learning strategy accelerated by the enhanced sampling method metadynamics (WTMetaD), which have already been successfully used to study several complex systems~\cite{bonati2018,niugallium2020,yang2022,  PhysRevB.107.064103}. The whole procedure of exploring the relevant reference configurations in the training set is illustrated in Fig.~\ref{F:NN_train}a. 

 First, we built an initial training set consisting of about 4000 configurations that were collected from a few short \textit{ab initio} MD (AIMD) trajectories. For AIMD calculations, two different CdSe crystals (hexagonal WZ and cubic ZB) were heated from 300 K to 2000 K into liquid.
 
Then, starting with these configurations and the corresponding DFT energies and atomic forces, and virials, we followed the active learning technique to extend and refine the training set. Each active learning iteration consists of the following four steps: (1) training 4 DNN potentials based on the same updated training set but with different initial training seeds; (2) exploration of the configurational space via DNN-based molecular dynamics (DPMD) simulations and their variants accelerated by WTMetaD (DP-WTMetaD). For these simulations, we performed different simulations using one of the 4 DNN potentials to sample the configurational space of different CdSe phases, such as hexagonal WZ, cubic ZB, stacking disordered, liquid, liquid-WZ, liquid-ZB, and activity transitions between these phases (see Fig.~\ref{F:NN_train}b). (3) selection of a small set of new relevant configurations for training set; (4) calculation of DFT energies, atomic forces, and virials for selected configurations. 
 
 For the selection of new candidate configurations in step (3), we chose the criterion of model deviation $\sigma$. The $\sigma$ is defined as the maximum standard deviation of the atomic forces predicted by four DNN potentials, $\sigma=\mathop{max}\limits_{x}\sqrt{\frac{1}{4}\sum_{\alpha=1}^{4}{\Vert{ \boldsymbol{F}_{x}^{\alpha}-\overline{\boldsymbol{F}_{x}}}\Vert^2}}$, where $\boldsymbol{F}_{x}^{\alpha}$ is the vector of atomic force on the atom $x$ predicted by the DNN potential $\alpha$, and $\overline{\boldsymbol{F}_{x}}$ is the average force on the atom $x$ over the four DNN potentials. We assumed that configurations with $\sigma$ lesser than ${\sigma}_{low}$ = 0.1 eV/{\AA} have already well-represented in the training set, while those with $\sigma$ larger than ${\sigma}_{high}$ = 0.45  eV/{\AA} are nonphysical and thus will be discarded. By this means, only structures with $\sigma$ in the range of [ 0.1, 0.45 ] eV/{\AA} were labeled as candidate configurations. The value of ${\sigma}_{low}$ and ${\sigma}_{high}$ were set up according to the suggestions given in Ref~\cite{zhang_dp-gen_2020}.  
 
 Finally, this active learning process will be exited when less than 10$\%$ candidate configurations were detected, as done in Ref~\cite{yang2021}, and a converged training set composed of about 24500 structures was obtained. The distribution of the configurational space of the training set is given in Fig.~\ref{F:NN_train}b. With the converged training set, we trained the final DNN potential.

All AIMD and DFT calculations were computed using the SCAN functional~\cite{Sun2015} as implemented in VASP~\cite{Kresse1993, Kresse1996}. We performed all DNN-based MD simulations using the LAMMPS~\cite{plimpton_fast_1995,thompson2022}  interfaced with the DeePMD-kit package, which was also used for training the DNN potentials. Enhanced sampling technique was provided by PLUMED~\cite{tribello_plumed_2014}. Additional computational details can be found in Supporting Information (SI). 
       
\section{Results and Discussion} 
\subsection{Validation of DNN potential}
\begin{figure}[!ht]
  \centering
  \includegraphics[width=0.9\columnwidth]{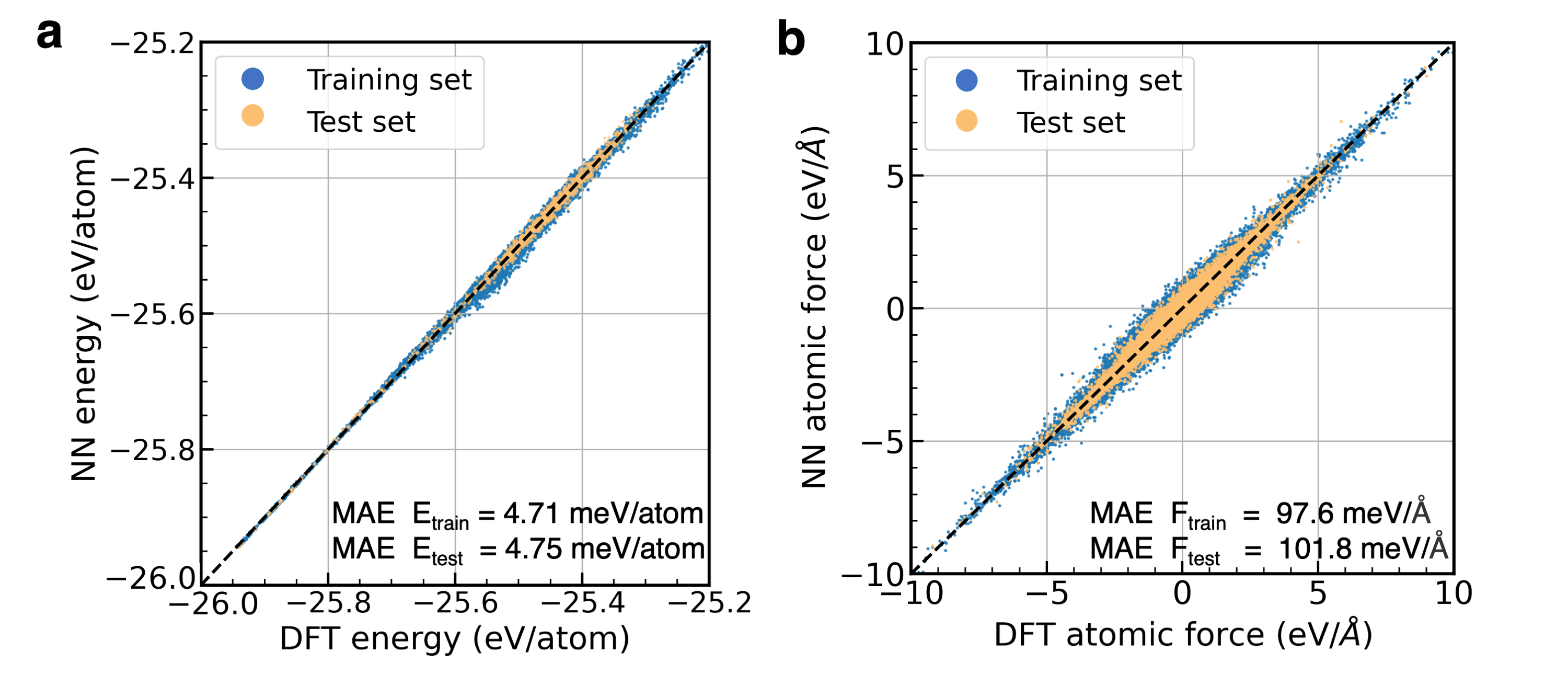}
  \caption{ Performance of the DNN potential. Comparison of (a) energies and (b) atomic forces between DFT and the final DNN potential for training set (blue dots) and test set (yellow dots). Black dashed lines are guides for perfect matches. 
  }
  \label{F:validation}
\end{figure}

To evaluate the performance of the final DNN potential, we first checked its ability to reproduce DFT results. The comparisons of energies and atomic forces calculated on training set and test set using DFT and DNN model are given in Fig.~\ref{F:validation}. The mean absolute errors (MAEs) of energies on the training and test sets are 4.71 meV/atom and 4.75 meV/atom, respectively. The MAEs of atomic forces on training and test sets are 97.6 meV/{\AA}  and 101.8 meV/{\AA}. The test set consists of 1500 configurations (See Fig. S5 for its configuration distribution in the 2D PCA space) collected from DPMD and DP-WTMetaD trajectories that were performed with the final DNN potential. In addition, the ground state energy of the cubic phase calculated with DNN potential is 0.6 meV/atom lower than that of the hexagonal phase, which is close to the DFT result of 0.9 meV/atom, indicating that the subtle energy difference between the cubic ZB and hexagonal WZ phases is well captured.

 Furthermore, we calculated the melting temperature of CdSe crystals in hexagonal WZ and cubic ZB phases using two-phase coexisting simulations. The potential energy evolution of the simulated trajectories for melting point calculation can be seen in Fig. S6. The melting temperature of hexagonal WZ phase was estimated to be 1465.0 $\pm$ 5.0 K, which is in good agreement with the experimental value of 1513.0 K~\cite{Lide2005}. The melting temperature of cubic ZB phase was estimated to be 1452.5 $\pm$ 2.5 K, 12.5 K
 lower than that of the hexagonal WZ phase. This result coincides with the experimental findings~\cite{fedorov1991determination} that the hexagonal WZ phase is more stable than the cubic ZB phase at high temperature.

\subsection{Brute-force MD simulation of homogeneous nucleation of CdSe}

\begin{figure}[!ht]
  \centering
  \vspace{0cm}
  \includegraphics[width=0.95\columnwidth]{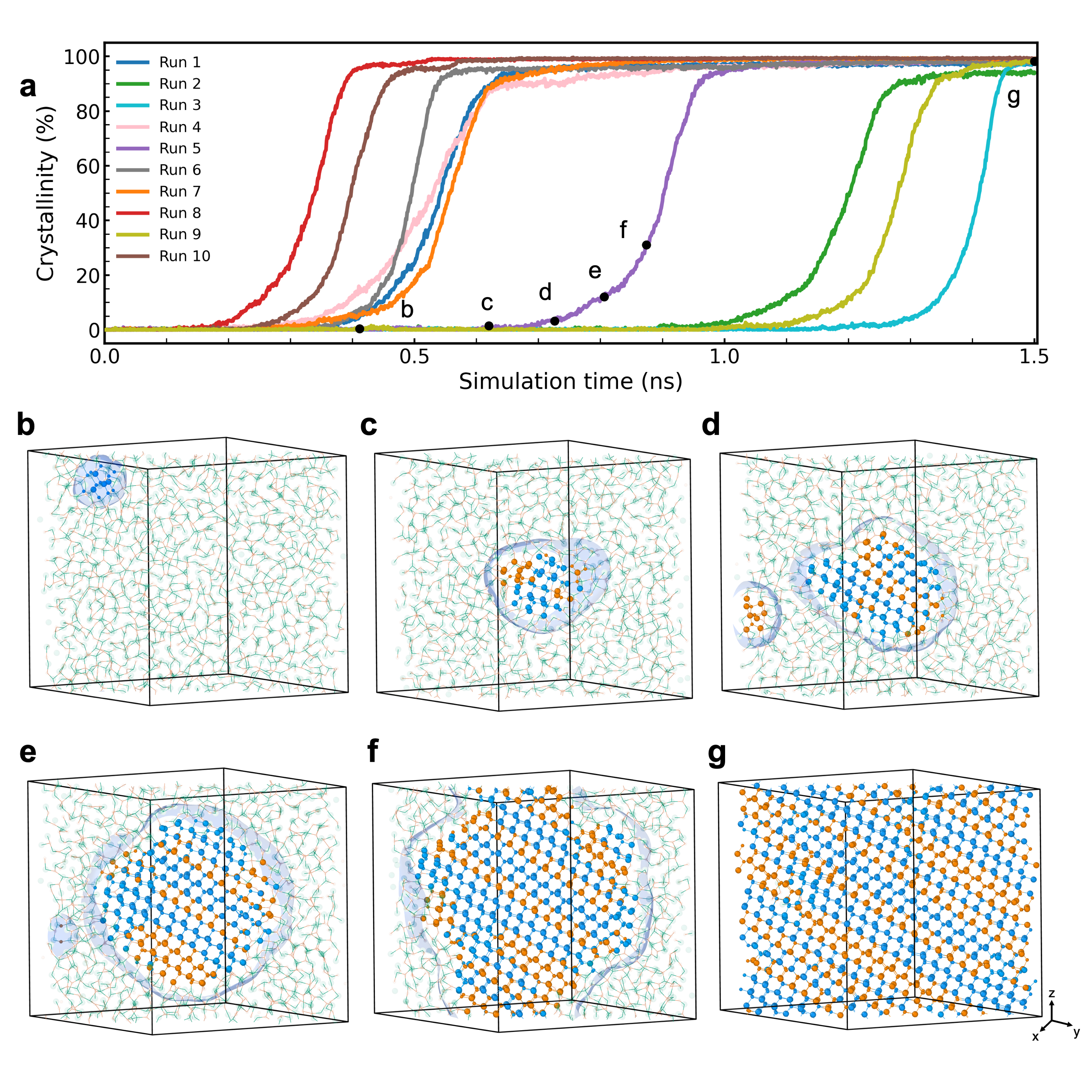}
  \caption{Homogeneous nucleation process of CdSe at 1000 K. (a) Percentage of solid-like atoms as a function of time for 10 independent trajectories simulated at 1000 K; Typical snapshots from one of the nucleation trajectories (Run 5 in (a)) at (b) 0.43 ns, (c) 0.62 ns, (d) 0.73 ns, (e) 0.81 ns, (f) 0.87 ns and (g) 1.50 ns, respectively. Orange and blue balls represent the CdSe crystals in hexagonal WZ and cubic ZB phases, respectively. The liquid phase of CdSe is indicated by light-colored lines. The blue transparent mesh represents the interface between the liquid and solid-like phases.}
  \label{F:nucleation}
\end{figure}

Upon cooling, liquid CdSe crystallizes into its solid phase. To investigate the evolution of the atomic structure of the nucleation process from scratch, we chose to perform brute-force MD simulations, ensuring that the intrinsic kinetics of the process were not altered. However, simulating homogeneous nucleation by MD is usually unachievable due to the large energy barrier that separates the liquid and solid phases. Thanks to the high efficiency of the DNN potential and deep supercooling condition we employed, we were able to obtain homogeneous nucleation trajectories from liquid CdSe. Fig.~\ref{F:nucleation} shows the evolution of the proportion of solid-like atoms throughout ten trajectories simulated at 1000 K. After periods of apparent but divergent incubation times, ten trajectories all crystallized. The mean incubation time was estimated to be 0.54 ns. Selected representative snapshots from the trajectory of the fifth run (labeled with lowercase letters) are presented in Fig.~\ref{F:nucleation}b-g, showing the homogeneous nucleation process of CdSe. During the incubation period, we noticed that multiple small solid-like nuclei emerged (Fig. S7) and disappeared (Fig.~\ref{F:nucleation}b-c). Stochastically, one of these nuclei survives and evolves into a larger nucleus (Fig.~\ref{F:nucleation}c-g), eventually causing the entire liquid phase to transform into the solid phase. During the growth process of the nucleus, other small nuclei can also be formed in the remaining liquid region of the system, as demonstrated in Fig.~\ref{F:nucleation}d. Such nuclei can disappear or merge into the dominant nucleus. 

The observation of the nucleation process shows apparently that the nucleus has a spherical-like shape, which minimizes the interfacial energy between solid and liquid phases. This phenomenon is similar to the nucleation process of silicon and ice~\cite{bonati2018, NiuIce2019}, suggesting that the classical nucleation theory (CNT) can approximately describe it. According to CNT, one can expect that the nucleation barrier decreases/increases as the temperature lowers/increases. To check this hypothesis, we also performed brute-force MD simulations at 900 K and 1100 K. These results indicated that multiple nuclei were formed in a relatively short time at 900 K (less than 1 ns), resulting in a polycrystalline structure (Fig. S8). In contrast, no successful nucleation event was observed at 1100 K within 25 ns. Fig.~\ref{F:nucleation} shows that the nucleus is made up of hexagonal WZ and cubic ZB structure layers, resulting in a stacking disordered structure, which is consistent with previous experimental findings~\cite{yoshiasa_mean-square_1997}. Interestingly, as the nucleus grows larger, the cubic ZB structure (colored blue) and hexagonal WZ structure (colored orange) can coexist in a single layer (see Fig.~\ref{F:nucleation}e-g). This suggests that the cubic and hexagonal structures compete with each other during the phase solidification. To gain a deeper understanding of how the stacking disordered structure is formed and the roles of Cd and Se atoms during this process, further investigations are conducted as follows.

\subsection{Temperature dependence of stacking disordered structure formation}
\begin{figure}
  \centering
  \includegraphics[width=0.95\columnwidth]{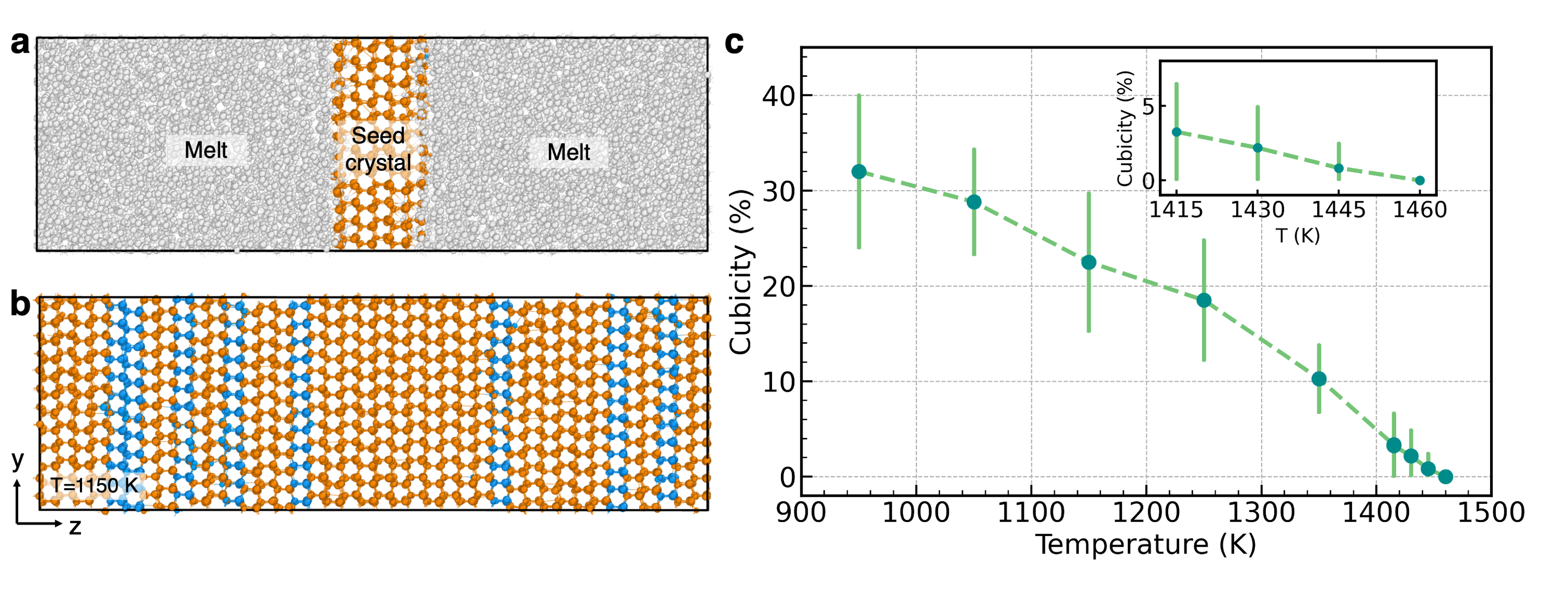}
  \caption{Temperature dependence of the formation of stacking disordered structure. (a) The initial solid-liquid simulation cell for crystal growth of CdSe; (b) One of the representative snapshots formed in the simulated crystal growth of CdSe at 1150 K; (c) The formed cubicity (fraction of cubic ZB stacking sequence) as a function of temperature. The error bars on cubicity were calculated from the standard deviation of 25 independent simulations at each temperature. Hexagonal WZ, cubic ZB, and melt (liquid) structures are colored yellow, blue and gray, respectively.}
  \label{F:cubicity}
\end{figure}

As discussed above, the relative stability and melting temperature of the cubic ZB phase differ from those of the hexagonal WZ phase, implying that the formation of the stacking disordered structure can be temperature dependent. To uncover the correlation between temperature and the formation of stacking disordered structures, we performed molecular dynamics simulations of the crystal growth of CdSe from its melt. Different thermodynamic points in the temperature range from 950 K to 1500 K (See Fig.~\ref{F:cubicity}) with a constant pressure of 1 bar were considered. For each temperature, we ran 25 independent simulations. These calculations were started from the same initial solid-liquid simulation cell consisting of 8640 atoms, where four hexagonal WZ layers of crystal seeds were included (Fig.~\ref{F:cubicity}a, see SI for more details). The growth direction was set to the \textit{z}-axis of the coordinate system (along the [001] direction of the hexagonal WZ phase).

To quantify the characteristics of the stacking disordered structure obtained from growth, we calculated the cubicity, which represents the fractions of cubic ZB stacking sequences~\cite{NiuIce2019,lupi2017}.
Fig.~\ref{F:cubicity}c shows the obtained temperature dependence of the cubicity. It is clear that the lower the temperature, the higher the value of cubicity. In other words, in the lower temperature region (\textless 1430 K), the initial solid-liquid mixed CdSe is more likely to grow into a stacking disordered structure (see Fig.~\ref{F:cubicity}b). Moreover, as we increase the temperature, the proportion of cubic ZB structure layers gradually decreases to zero. For instance, at 1430 K, 14 out of 25 trajectories yielded a pure hexagonal WZ phase, while the average cubicity of all 25 trajectories is only approximately 2\%. Furthermore, when we increased the temperature to 1460 K, slightly lower than the melting temperature of the hexagonal phase (T$_m$ = 1465 $\pm$ 5 K), no stacking disordered structure was observed. These results suggest that 1430 K (corresponding to 35 K of supercooling) is approximately the lowest temperature at which we could obtain a pure hexagonal WZ phase, which is in line with the experimental Bridgman method where the low-temperature region is 50 K below the melting temperature~\cite{ni_growth_2018}. This temperature dependence of crystallization explains why the experimental temperature to obtain the pure hexagonal WZ phase with the Bridgman method should be kept not lower than 50 K from the melting point.

\subsection{Structural competition in crystallization process}
   \begin{figure}[!ht]
   \vspace{0cm} 
  \centering
  \includegraphics[width=0.95\columnwidth]{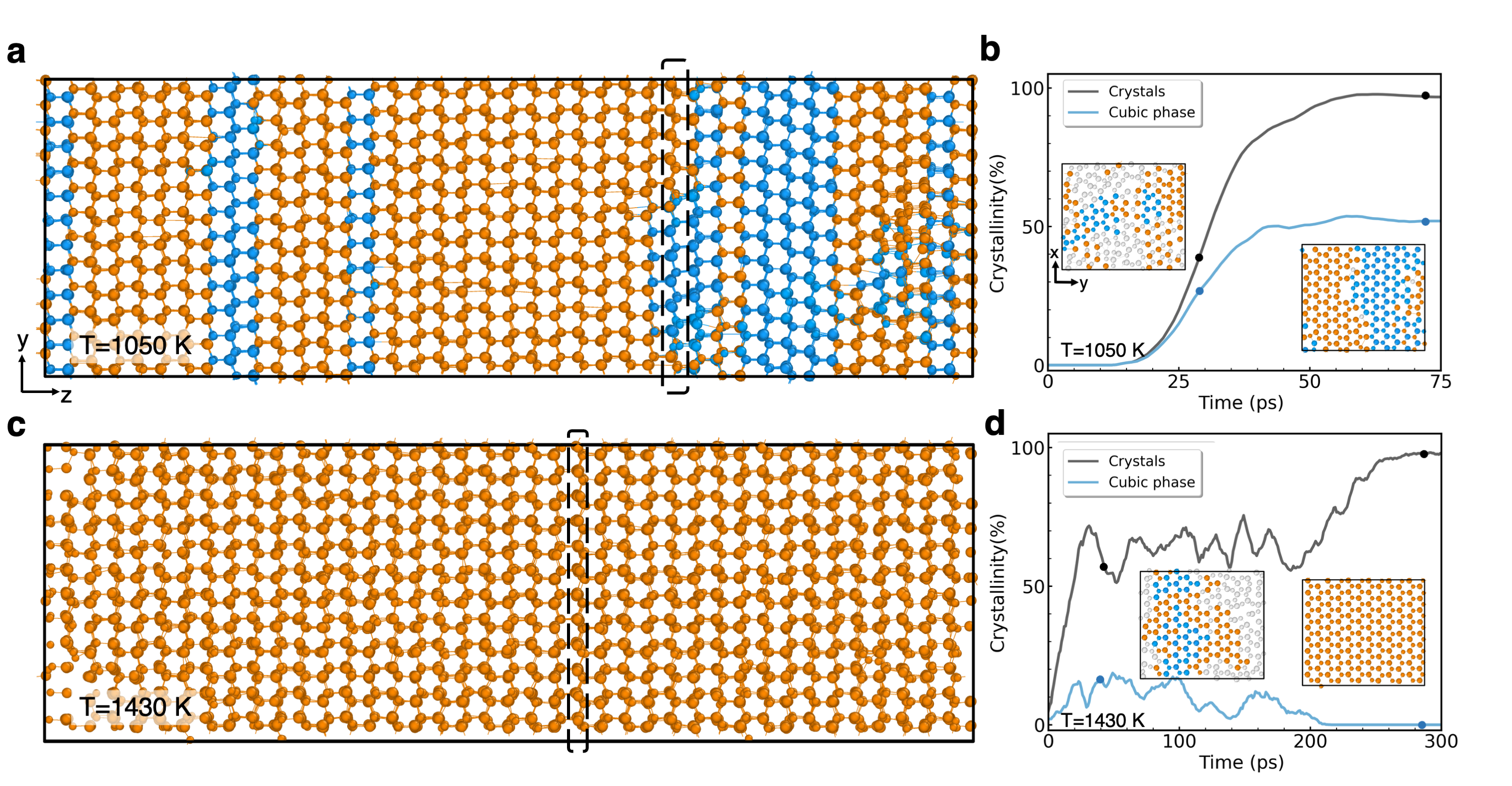}
  \caption{
  Illustration of structural competition during the crystal growth process. Snapshots of two representative crystals obtained in the growth simulation at (a) 1050 K and (c) 1430 K, respectively; (b) and (d) show the solid-like evolution of atoms from two atomic layers of the black dashed box in (a) and (c) along with the CdSe growth progress, and some typical snapshots of the instantaneous local magnifications (projected in the \textit{x-y} plane) of these atomic layers. The crystalline phases of hexagonal WZ, cubic ZB, and liquid structures are colored yellow, blue and gray, respectively. 
  }
  \label{F:sc}
\end{figure}

It is interesting to understand why it is easier to obtain a pure hexagonal phase when we increase the temperature of crystal growth. Actually, during the crystal growth process, various defects appeared and persisted, and their performances were closely related to temperature. Two crystal snapshots (with a projection of the \textit{y-z} plane) taken from typical crystallization simulations at 1050 K and 1430 K are illustrated in Fig.~\ref{F:sc}a and c, respectively. For each case, we followed the crystal formation of two atomic layers highlighted by the black box, as shown in Fig.~\ref{F:sc}b and d (with a projection of the \textit{x-y} plane). It is obvious that in both situations, stacking competition between cubic ZB and hexagonal WZ in the same layer occurred during crystal growth. As the crystal grows, for the case of 1050 K, cubic ZB atoms are frozen without further rearrangement, resulting in a manner of two-phase coexistence (Fig.~\ref{F:sc}a). In contrast, for the case of 1430 K, cubic ZB atoms are more likely to be converted into the hexagonal WZ phase after a period of structural fluctuation and rearrangement, and the proportion of cubic ZB atoms fluctuates and 
approaches zero as time evolves, leading to a pure phase of hexagonal WZ (Fig.~\ref{F:sc}c).

\subsection{Microscopic mechanism of CdSe crystallization}
\begin{figure}[!ht]
  \centering
  \includegraphics[width=1\columnwidth]{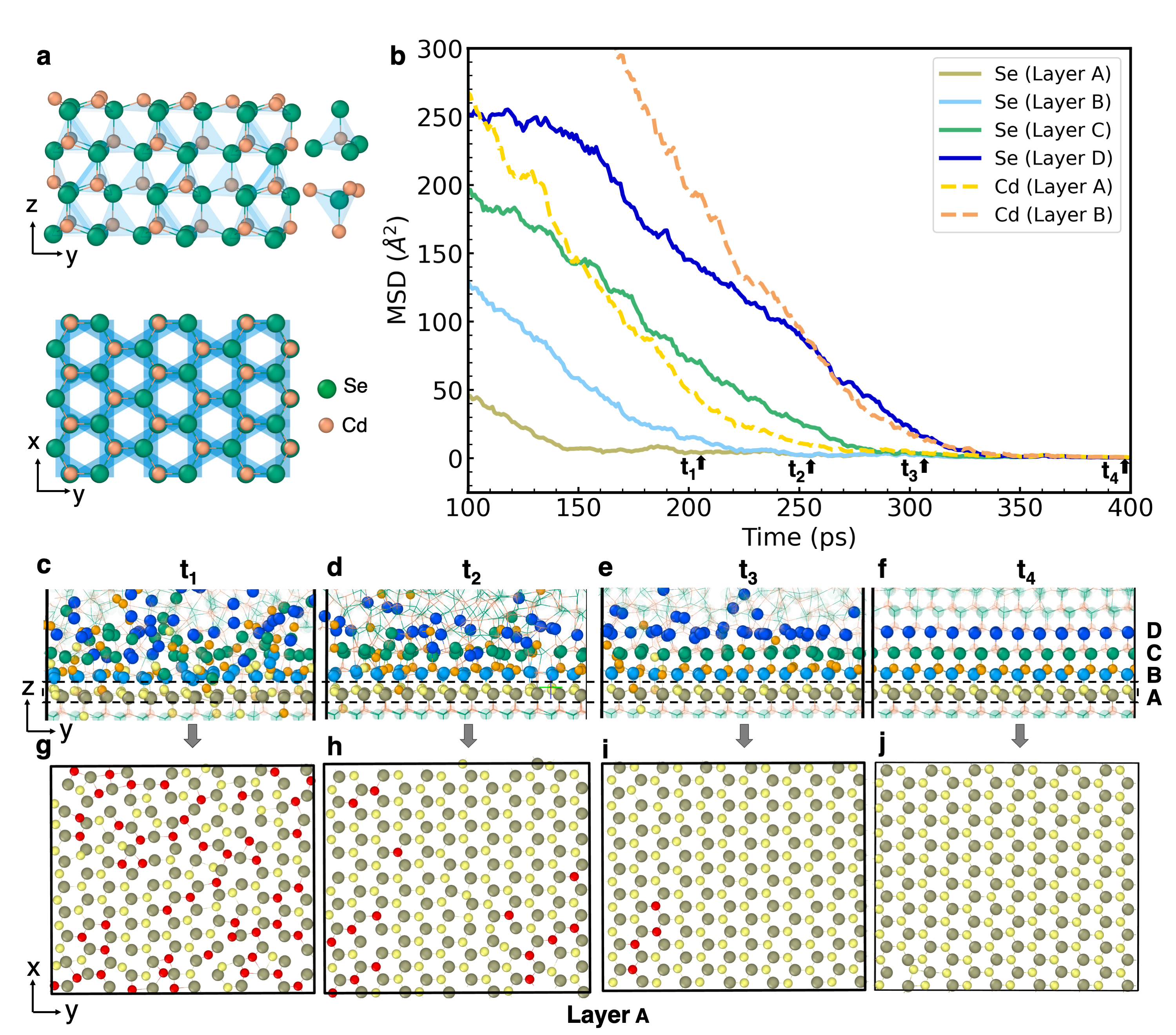}
  \caption{MSD analysis and snapshots from the crystallization process of CdSe layers. (a) Crystal structures of hexagonal WZ CdSe projected in the \textit{y-z} and \textit{x-y} planes. (b) The MSDs for Cd and Se atoms in different layers as a function of time. (c-f) Typical snapshots in the  \textit{y-z} plane for four CdSe layers (labeled with capital letters of \textbf{A}, \textbf{B}, \textbf{C} and \textbf{D}) from the crystallization process (t$_1 $to t$_4$). (g-j) Instantaneous local
magnifications (projected in \textit{x-y} plane) of atoms in layer \textbf{A}. Se atoms in layers \textbf{A}, \textbf{B}, \textbf{C} and \textbf{D} are colored grey, light blue, green, and dark blue, respectively. Cd atoms in layers \textbf{A} and \textbf{B} are colored yellow and orange, respectively. Notably, the Cd atoms in layer A that contributed to the MSD value were highlighted in red.
  }
  \label{F:msd}
\end{figure}

The structure of solid CdSe consists of an equal number of tetracoordinated Cd and Se atoms placed at alternate tetrahedral points of a diamond lattice, as depicted in Fig.~\ref{F:msd}a. A straightforward hypothesis of its crystallization process would be that each type of atom is attached to the solid-liquid interface one by one so that an alternating arrangement of Cd and Se atoms can be formed. Nevertheless, a careful examination of the 
 the evolution process of the interfacial region during the growth process reveals a completely different narrative, where 
the behaviors of Cd and Se atoms exhibits significant disparties.

The detailed crystallization mechanism of CdSe was studied using one of the representative trajectories of the CdSe crystal growth at 1415 K. Specifically, we followed the formation processes of four new 
crystallized  layers denoted by capital letters \textbf{A}, \textbf{B}, \textbf{C}, and \textbf{D} in Fig.~\ref{F:msd}f, respectively. To quantify the dynamical behaviour of each type of atom from different layers, we calculated the mean square displacements (MSDs) for Cd and Se atoms in each layer as a function of time. The MSD value for atoms in each layer at time \textit{t} was calculated as follows and the instantaneous structure at time $t_4 = 400$ ps when the studied four atomic layers crystallized was set as a reference point.
\begin{equation}
\textbf{MSD}_{(t)} = \frac{1}{N}\sum_{i=1}^{N}{<|\textbf{r}_i\left(t\right)-\textbf{r}_i\left(t_4\right)|^2>}\,
\end{equation}
in which \textit{N} means the number of Cd or Se atoms in each layer and \textbf{r}$_i$(\textit{t}) is the position of the \textit{i$^{th}$} atom at time \textit{t}.

The obtained MSDs for Cd and Se atoms in different layers as a function of time are presented in Fig.~\ref{F:msd}b. The smaller the MSD value, the higher the crystallinity of the atomic layer. Once all Cd or Se atoms in a given layer were crystallized, the corresponding MSD curves approach zero. 
Counterintuitively, we find that the MSDs of Cd and Se atoms for the same layer do not reach zero simultaneously, rather an approximately 100 ps delay is noticed in the MSD curve of Cd compared with that of Se as shown in Fig.~\ref{F:msd}b, indicating a rather complex mechanism underlying the crystallization process of CdSe. 

For instance, at \textit{t$_1$} (see Fig.~\ref{F:msd}c and Fig.~\ref{F:msd}g), Se atoms in layer \textbf{A} were well arranged in their lattice with an MSD value close to zero (dark-khaki line in Fig.~\ref{F:msd}b), while the Cd atoms in the same layer still have a relatively large MSD value of 45 {\AA}$^2$ (yellow dotted line in Fig.~\ref{F:msd}b). 

Furthermore, we find that the crystallization of Cd atoms is not only affected by its short-range atomic environment but also strongly connected with the medium-range ordering of Se and Cd atoms. 
As illustrated in Fig.~\ref{F:msd}d at $t_2$, all lattice points of Se atoms in both layers \textbf{A} and \textbf{B} were filled, resulting in tetrahedrons that favor the occupation of the Cd atoms in their center. 

The cross-section snapshot of layer \textbf{A} (Fig.~\ref{F:msd}h) confirmed this argument that all tetrahedron centers of Se clusters were placed with Cd atoms. However, some of these Cd atoms (red marked atoms in Fig.~\ref{F:msd}h) were not yet stable and would migrate to other places as the crystal grew. With further analysis, we found that the crystallization of Se atoms in layer \textbf{C} was essential to stabilize the Cd atoms in layer \textbf{A}. Once the lattice sites of Se atoms in layer \textbf{C} were arranged, similarly, tetrahedral sites consisting of Se atoms from layer \textbf{B} and \textbf{C} were formed in favor of the occupation of the Cd atoms in layer \textbf{B}. Importantly, this arrangement stabilized the Cd atoms in layer \textbf{A} by forming perfect tetrahedrons consisting of Cd atoms from layers \textbf{A} and \textbf{B}. 
In addition, the MSDs of Cd atoms in layer A and Se atoms in layer C reach zero approximately concertedly at \textit{t$_3$}. 

A similar phenomenon was also observed in layer \textbf{B}, in which Cd atoms in this layer were stabilized until Se atoms in layer \textbf{D} found their positions.
Therefore, the cooperative actions between Cd and Se atoms within the medium range region dominate the crystallization process of CdSe.

We suggest that this quite different behavior of Se and Cd atoms during the crystallization process is closely related to the radii of Se$^{2-}$ and Cd$^{2+}$ ions. The ionic radius of Se$^{2-}$ is 1.84 {\AA}, while that of Cd$^{2+}$  is only 0.92 {\AA}~\cite{PARASYUK20021}. This large difference in ion radii makes Cd atoms more mobile than Se atoms. To confirm this argument, the diffusion coefficients $D$ for Se$^{2-}$ ({$D_{Se^{2-}}$}) and Cd$^{2+}$({$D_{Cd^{2+}}$}) at temperatures from 1100 K to 1500 K for liquid CdSe (See SI for more details) were calculated. We found that the calculated {$D_{Cd^{2+}}$} were at least two times larger than {$D_{Se^{2-}}$}. 

\begin{figure}[!ht]
  \centering
  \includegraphics[width=1\columnwidth]{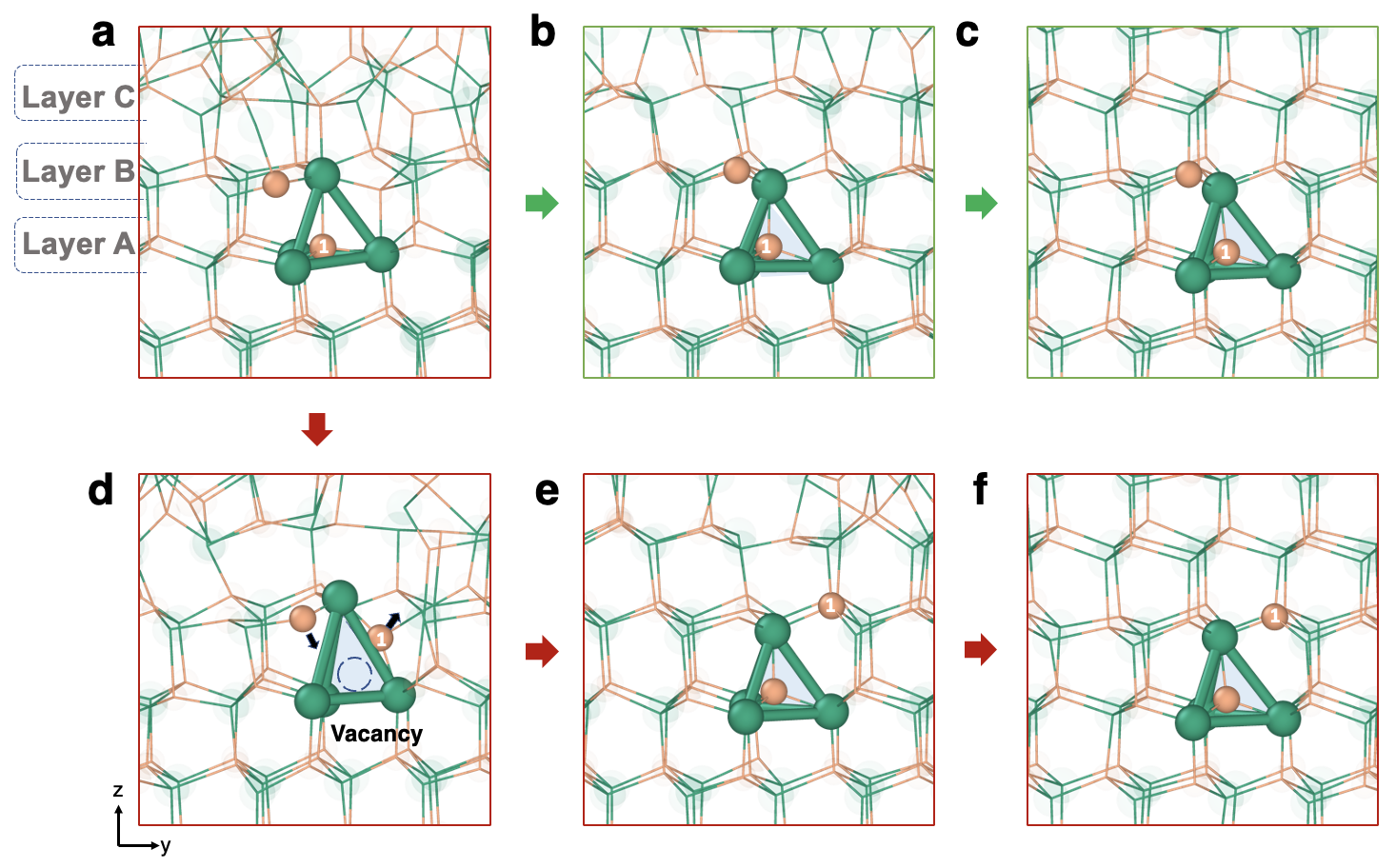}
  \caption{Schematic diagram of the microscopic mechanism of CdSe growth. Two possible pathways: a $\rightarrow$ b $\rightarrow$ c (green arrow marked) and a $\rightarrow$ d $\rightarrow$ e $\rightarrow$ f (red arrow marked) are  illustrated. The green tetrahedron cluster is composed of Se atoms from layers \textbf{A} and \textbf{B}, and Cd atoms are colored brown.
   }
  \label{F:mechanism}
\end{figure}

To gain a closer mechanism of the crystallization of Cd atoms, we analysed the movement of Cd atoms in layer \textbf{A} located in the tetrahedral center of Se, corresponding to those discussed in Fig.~\ref{F:msd}h at time $t_2$. Two possible pathways as illustrated in Fig.~\ref{F:mechanism} were located. Cd atoms can either stay at the tetrahedron center of Se (pathway: a $\rightarrow$ b $\rightarrow$ c) or migrated to other layers (pathway: a $\rightarrow$ d $\rightarrow$ e $\rightarrow$ f). At the starting point, Se atoms in layers \textbf{A} and \textbf{B} were well arranged at their lattice sites forming a perfect tetrahedron (green ball-and-stick marked in Fig.~\ref{F:mechanism}a), while those in layer \textbf{C} were randomly distributed. Stabilization by the tetrahedron of Se, Cd atoms preferred to be located at the center of the tetrahedron resulting in a pathway of a $\rightarrow$ b $\rightarrow$ c. However, without the stabilization from Se atoms in layer \textbf{C}, Cd atoms in layer \textbf{B} are super mobile, which pushes the Cd$_1$ atom in layer \textbf{A}  to move away from the original site and leave a vacancy (Fig.~\ref{F:mechanism}d). Subsequently, Cd atoms in layer \textbf{B} moved to this vacancy and occupied the original position of Cd$_1$ (Fig.~\ref{F:mechanism}d-e), leading to another pathway of a $\rightarrow$ d $\rightarrow$ e $\rightarrow$ f. Taken together, we can see that Cd atoms in layer \textbf{A} were not stabilized until their third-shell neighbor atoms Se find their position. Therefore, the medium-range ordering of Se atoms governs the nucleation and growth process of CdSe. 

\section{Conclusion}
In this work, we have developed a DNN potential with \textit{ab initio} accuracy using an enhanced sampling-accelerated active learning approach to investigate
the crystallization process of CdSe. With this potential, we conducted molecular dynamics simulations for systems containing
8640 atoms, enabling us to systematically study the crystallization mechanism of CdSe at atomic level. Brute-force simulations were first performed to study the nucleation process of CdSe.  We found that a spherical-like nucleus was formed, consistent with classical nucleation theory. To elucidate the growth mechanism, we then performed MD simulations of CdSe growth from its melt at temperatures ranging from 950 K to 1500 K. Our results demonstrated that the formation of stacking disordered structures was temperature-dependent, and a pure hexagonal WZ crystal can only be obtained above 1430 K, 35 K below its melting temperature, in line with the experimental observations using the Bridgman method. By analysing the crystallization process in detail, we found that the dynamic behaviors and solidification paces of Cd and Se atoms were markedly different, primarily due to the difference between the ionic radius of Se$^{2-}$ and Cd$^{2+}$. Specifically, lower mobility Se atoms pioneered the solidification process by forming tetrahedrons, followed by Cd atoms occupying centers of these tetrahedrons and settling down until the third-shell neighbor of Se atoms found their lattice positions. In other words, the medium-range ordering of Se atoms governs the crystallization process of CdSe. Consequently, the underlying crystallization mechanism of CdSe, shaped by the distinct dynamical behavior of Cd and Se atoms,  is much more intricate than anticipated.
Our findings highlight the crucial role of the complex dynamical process that governs the quantity of synthesized crystals.  Molecular dynamics simulation aided by $ab$ $initio$ quality machine learning potential is a powerful tool for studying such processes.

\section*{CRediT authorship contribution statement}
\textbf{Linshuang Zhang:} Performed calculations, Analyzed the data, Wrote the paper.
\textbf{Manyi Yang:} Conceptualization, Methodology, Performed calculations, Analyzed the data, Wrote the paper.
\textbf{Shiwei Zhang:} Performed enhanced sampling simulations, Analyzed the data.
\textbf{Haiyang Niu}: Conceptualization, Methodology, Analyzed the data, Wrote the paper, Funding acquisition, Supervision. 

\section*{Declaration of competing interest}
The authors declare that they have no known competing financial interests or personal relationships that could have appeared to influence the work reported in this paper.

\section*{Acknowledgment}
The authors would like to thank Mingfeng Liu, Jingwei Zhang, Junwei Hu and Mingyi Chen for their valuable contributions to the discussions.  This work was supported by the National Natural Science Foundation of China (grant No. 22003050), the National Science Fund for Excellent Young Scientist Fund Program (Overseas) of China, the Science and Technology Activities Fund for Overseas Researchers of Shaanxi Province, China, and the Research Fund of the State Key Laboratory of Solidification Proceeding (NWPU) of China (No. 2022-QZ-03). The calculations were supported by the International Center for Materials Discovery (ICMD) cluster of NWPU and the Swiss National Supercomputing Centre (CSCS) under project ID $S1134$ and $S1183$.

\bibliographystyle{elsarticle-num} 
\bibliography{Article}
\end{document}


\begin{frontmatter}
\biboptions{numbers,sort&compress}
\title{Supplementary material\\}
\title{\textit{Ab initio} investigation of the crystallization mechanisms of cadmium selenide}

\author[label1]{Linshuang Zhang}
\affiliation[label1]{organization={State Key Laboratory of Solidification Processing, International Center for Materials Discovery, School of Materials Science and Engineering,
Northwestern Polytechnical University},
            city={Xi'an},
            postcode={710072}, 
            country={China}}
\author[label2]{Manyi Yang\corref{cor1}}
\ead{manyi.yang@iit.it}
\affiliation[label2]{organization={Italian Institute of Technology},
            addressline={Via E. Melen 83}, 
            city={Genoa},
            postcode={16152}, 
            country={Italy}}
\author[label1]{Shiwei Zhang}           
\author[label1]{Haiyang Niu\corref{cor1}}
\ead{haiyang.niu@nwpu.edu.cn}
\cortext[cor1]{Corresponding author}

\end{frontmatter}

\clearpage

\section{Polymorphism of CdSe crystal}
\begin{figure}[!ht]
  \centering
  \includegraphics[width=0.9\columnwidth]{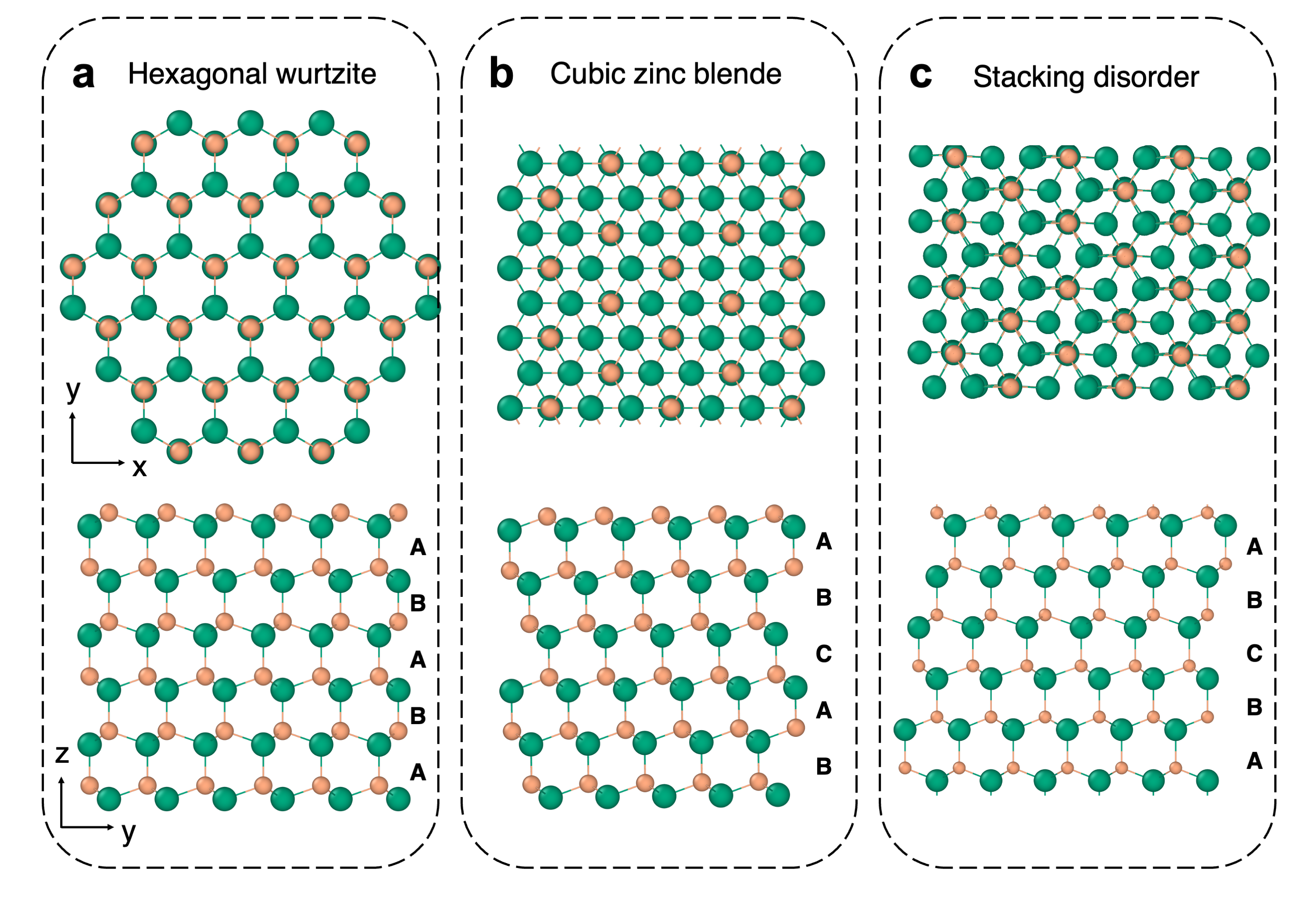}
  \caption{Schematic diagrams of \textit{x-y} and \textit{y-z} plane projections of (a) hexagonal wurtzite (WZ), (b) cubic zinc blende (ZB) and (c) stacking disordered phases of CdSe crystal. In all snapshots of this work, green and brown spheres refer to Se and Cd atoms, whose sizes are proportional to the radii of their corresponding ions, respectively.
  }
  \label{F:base}
  \end{figure}

\section{Computational details}
\subsection{AIMD simulations}
All AIMD simulations were performed using Vienna \textit{ab-initio} Simulation Package (VASP)~\cite{KRESSE199615}. The canonical ensemble (NVT) with periodic boundary conditions was used. The time step was set to 1 fs and the temperature was controlled using the stochastic velocity rescaling thermostat with a coupling constant of 40 fs. Energies and forces were computed using meta-GGA XC functional of strongly constrained and appropriately normed (SCAN)~\cite{Sun2015}. The energy convergences were set to 10$^{-5}$ eV. The Monkhorst-Pack \textit{k}-point gird was set to 1$\times$1$\times$1 and the plane-wave cutoff energy was set to 300 eV. In these AIMD calculations, two different phases of CdSe crystal (hexagonal WZ and cubic ZB phases) made of 96 atoms were heated from 300 K to 2000 K into liquid.
  
\subsection{DFT single point calculations}
\begin{figure}[!ht]
  \centering
  \includegraphics[width=1.0\columnwidth]{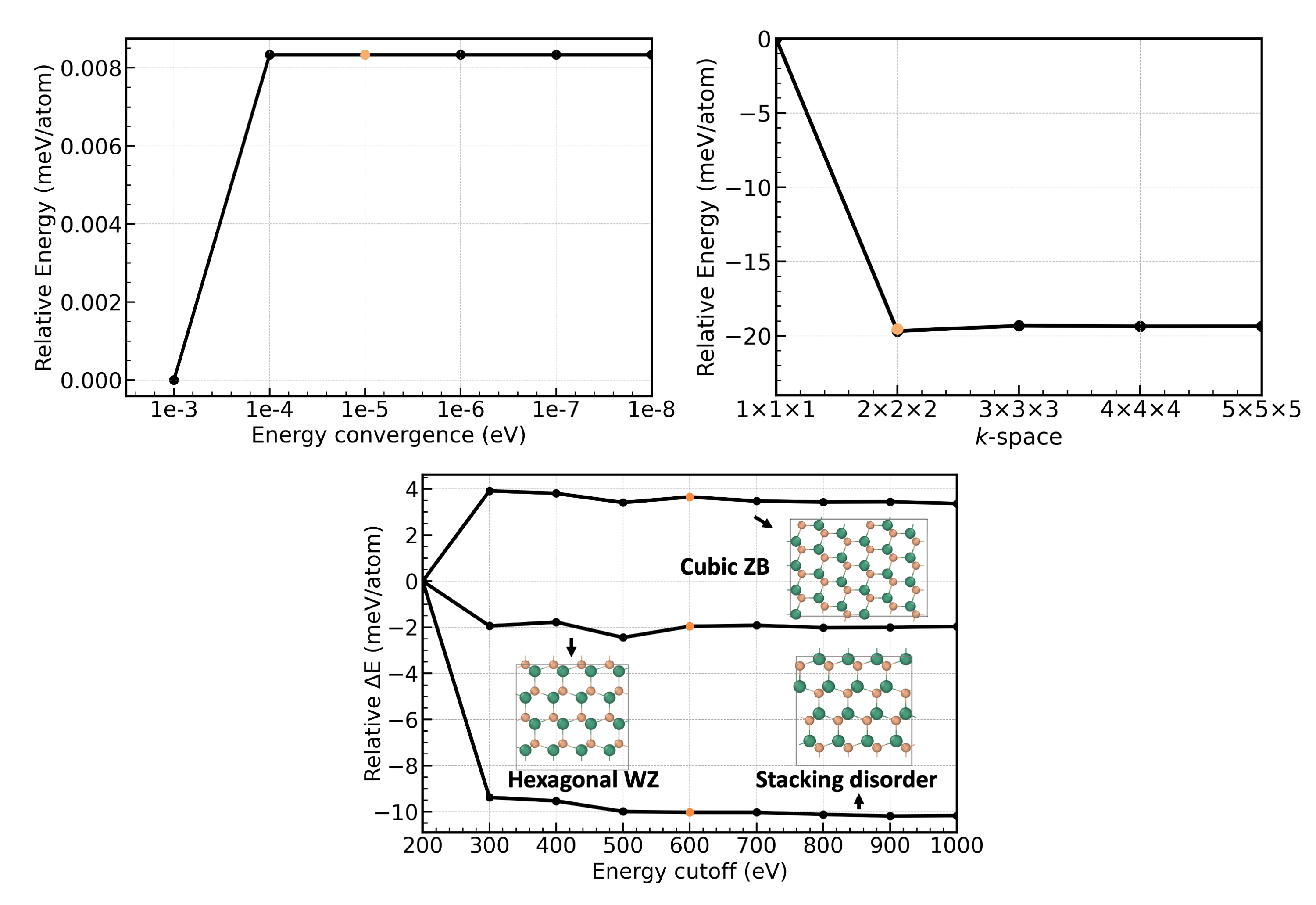}
  \caption{The convergence test for the threshold of energy convergence, \textit{k}-space and energy cutoff.}
  \label{F:benchmark}
\end{figure}

DFT single point energies, forces and virials needed for NN training were calculated with meta-GGA XC functional of SCAN~\cite{Sun2015}, as implemented in the package of VASP. The plane-wave cutoff was set to 600 eV. We used a 2$\times$2$\times$2 \textit{k}-point Monkhorst-Pack grids. The threshold of energy convergence was set to 10$^{-5}$ eV. These setups provide more accurate results as illustrated in Fig.~\ref{F:benchmark}.

\subsection{Collective variables}
 We chose the structure factor as the collective variable (CV) to drive the solid-liquid phase transition, which has been proven to be a powerful CV for studying the nucleation processes for different crystals~\cite{niusilica2018,NiuIce2019}. 
 Specifically, the CV is constructed as a combination of two descriptors $s$  = $s_{100}^{xy}+ s_{110}^{xy}$. The descriptor $s_{100}^{xy}$ and $s_{110}^{xy}$ corresponds to the intensities of the two main peaks of one single honeycomb layer which is projected into the \textit{x-y} plane. In such scenario, $s$ is blind with respect to the hexagonal WZ, cubic ZB and stacking disordered phases, while it can distinguish the solid phase from the liquid phase well (Fig. S3). The descriptor $s_{hkl}^{xy}$ can be calculated from the expression: 
\begin{equation}
s_{hkl}^{xy}(Q)=\frac{1}{N}\sum_{i=1}^{N}{\sum_{j=1}^{N}f_i(Q)f_j(Q)J_0(Q\cdot R_{ij}^{xy})\cdot \omega^{xy}(R_{ij}^{xy})\omega^{z}}(R_{ij}^z)\,,
\label{s-xy}
\end{equation}
where \textit{J$_0$} is the \textit{0$^{th}$} order of the first kind Bessel function, $hkl$ refers to the Miller indexes (100) and (110),   
$Q$ is the scattering vector, $f_i(Q)$ and $f_j(Q)$ are the atomic scattering factors, \textit{R$_{ij}^{xy}$} is the distance between atoms \textit{i} and \textit{j} in the \textit{x-y} plane. $\omega^{xy}(R_{ij}^{xy} )=\frac{\ 1\ }{1+e^{\sigma(R_{ij}^{xy}\ - R_c^{xy}\ )}}$ and $\omega^{z}(R_{ij}^{z} )=\frac{\ 1\ }{1+e^{\sigma(R_{ij}^{z}\ - R_c^{z}\ )}}$ refer to switching functions to make the descriptor smooth, where $R_c^{ij}$ and $R_c^{z}$ were set to 10 {\AA} and 2 {\AA} in this work, respectively. The wavelength $\lambda$ is set to 1.5406 \AA. 

\begin{figure}[!ht]
  \centering
  \includegraphics[width=0.9\columnwidth]{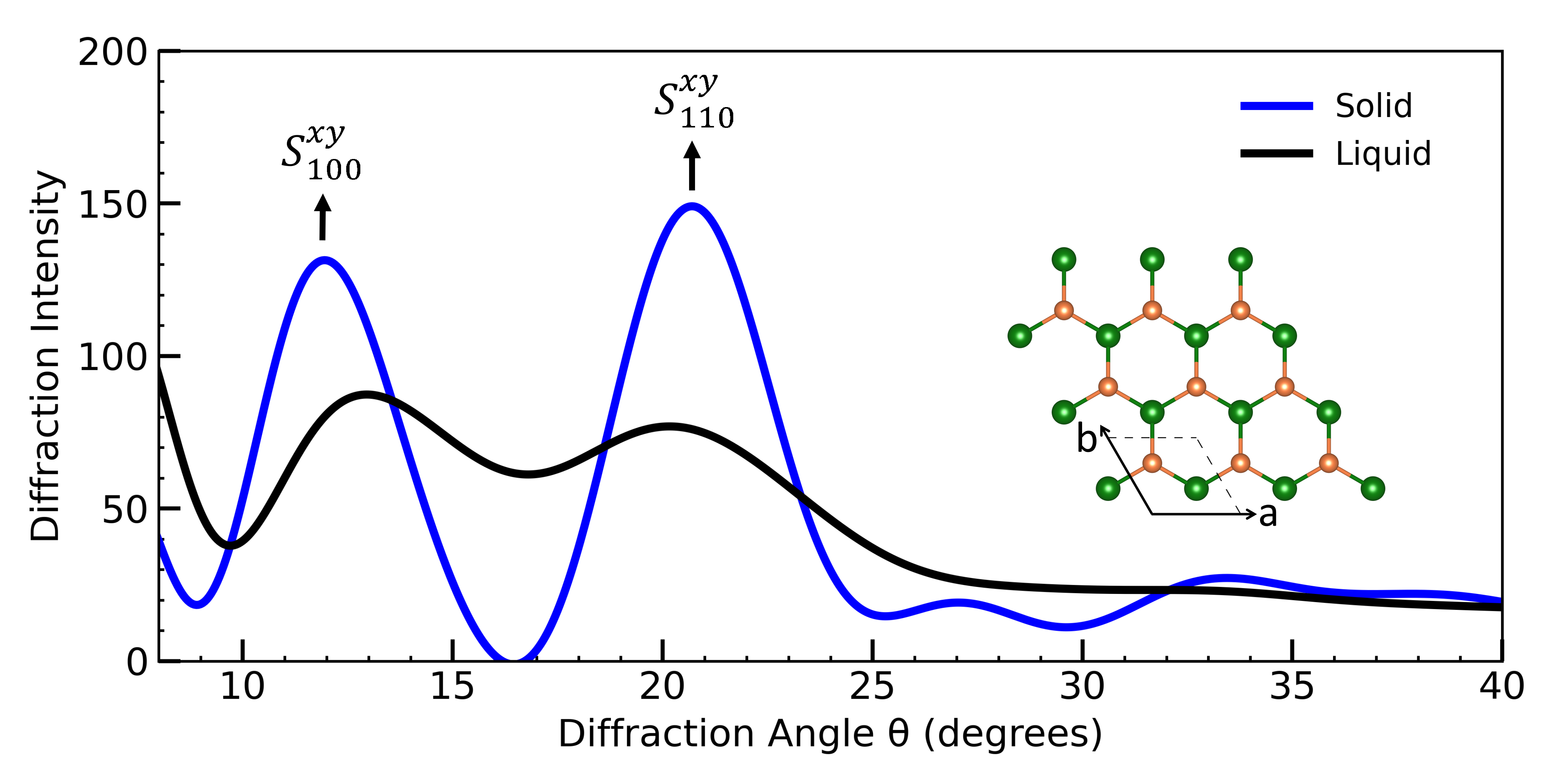}
  \caption{Simulated structure factors of the solid and liquid phases from the two-dimensional (2D) perspective. The components of CVs are highlighted with black arrows. The schematic of the \textit{x-y} plane for the solid phase is given in the inset. 
  }
  \label{F:xrd}
  \end{figure}

\subsection{DeepMD setup}
We used the Deep Potential-Smooth Edition scheme~\cite{NEURIPS2018_e2ad76f2} as implemented in the DeePMD-kit package~\cite{wang2018}, to build a deep neural network (DNN) potential to represent the free energy surface of CdSe. We used three hidden layers with (30, 60, 120) nodes/layer and a matrix size of 16 for the embedding network, and four hidden layers with 240 nodes/layer for the fitting network. ResNet architecture was used for both networks. The cutoff radius was set to $6.5$ {\AA} and the descriptors decay smoothly from $6.0$ {\AA} to $6.5$ {\AA}. The batch size was set to $8$. The hyper-parameters of \textit{start\_pref\_e}, \textit{start\_pref\_f}, \textit{start\_pref\_v}, \textit{limit\_pref\_e}, \textit{limit\_pref\_f} and \textit{limit\_pref\_v} in loss function were set to 0.02, 1000, 0.02, 2.0, 1.0, and 2.0, respectively. The learning rate exponentially decays from $1.0 \times 10^{-3}$ to $1.0 \times 10^{-8}$. A small training step of 4 × 10$^{5}$ was used for the active learning process, while for training the final DNN potential, this number was increased to 1 × 10$^{6}$. 

\subsection{DNN-based molecular dynamics}
All DNN-based molecular dynamics (DPMD) simulations were performed by patching the DeepMD-kit~\cite{wang2018} software implemented in LAMMPS~\cite{PLIMPTON19951} and PLUMED 2~\cite{TRIBELLO2014604}. The ensemble of NPT was used in all calculations with a time step of 1 fs and a pressure of 1 bar. Temperature and pressure were controlled using Nos\'e-Hoover thermostat~\cite{nose1984,nose1985} and  Nos\'e-Hoover-like barostat~\cite{barostat1994}with coupling constants of 0.1 ps and 1 ps, respectively. In order to prevent the uniform rectilinear motion of the whole system, the linear momentum of each atom was rescaled by subtracting the momentum of the center of mass every 10-time steps.

\section{Calculated model systems}
In this work, different simulations were performed on models of varying system sizes.

\begin{figure}[!ht]
  \centering
  \includegraphics[width=0.9\columnwidth]{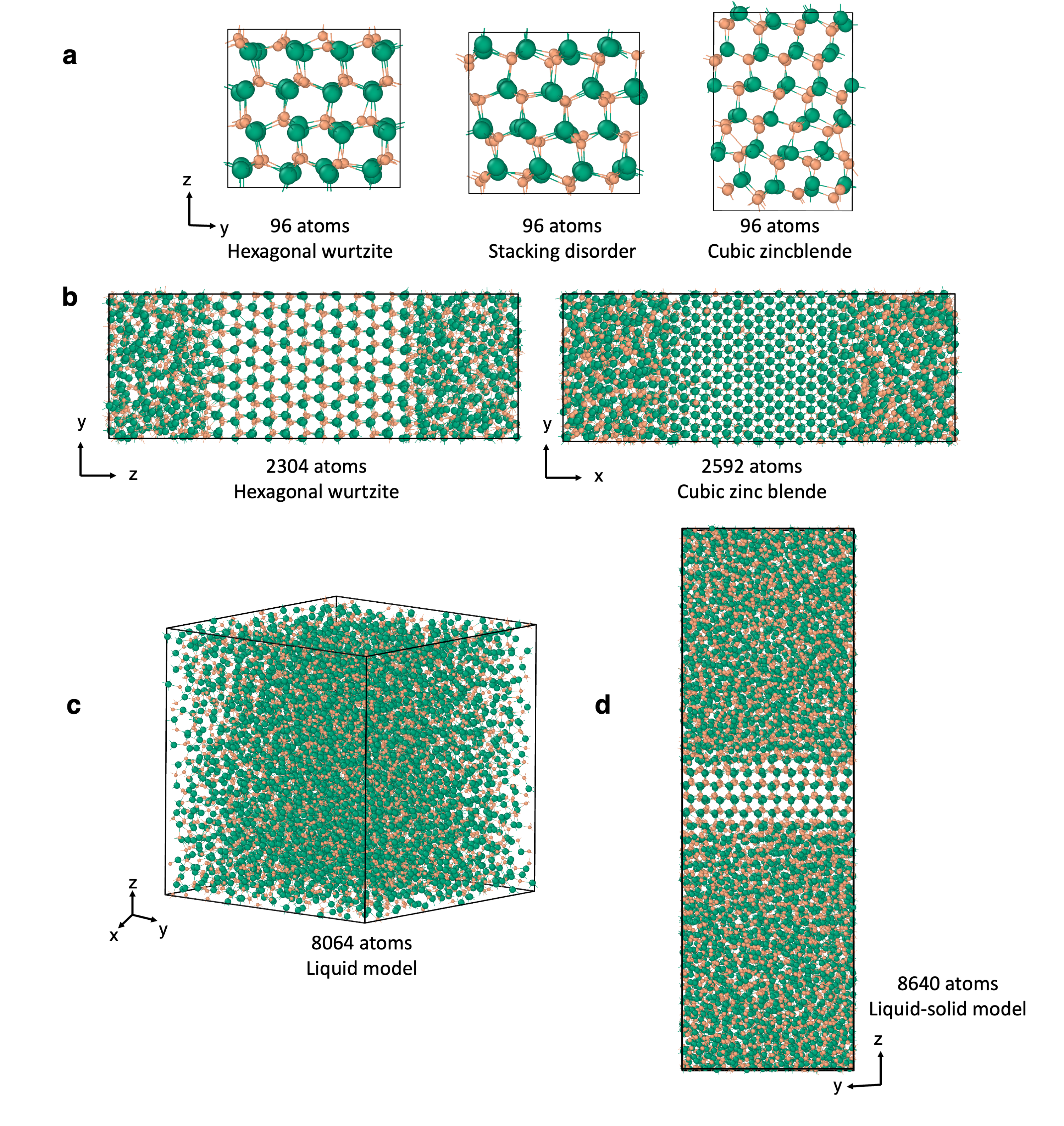}
  \caption{ (a) Three models used to collect configurations for training and test sets; (b) Two solid-liquid models to estimate the melting temperature of CdSe crystals in hexagonal WT and cubic ZB phases; (c) A liquid model used to study the homogeneous nucleation of CdSe; (d) Initial solid-liquid model used to study the growth of CdSe from its melt.}
  \label{F:initialmodel}
\end{figure}

For simulations of collecting configurations for training and test sets, small models made of 96 atoms, as shown in Fig.~\ref{F:initialmodel}a, were used, considering the expensive computational cost of DFT. 

To estimate the melting temperature of CdSe crystals, we constructed solid-liquid supercells made of 2304 and 2592 atoms, corresponding to hexagonal-liquid and cubic-liquid models with edges of 80.2, 30.4, 28.7 {\AA} and 76.7, 35.2, 32.2 {\AA}, respectively (see Fig.~\ref{F:initialmodel}b). These solid-liquid supercells were prepared according to the following three steps: First, crystal cells (hexagonal WT and cubic ZB) were relaxed at T = 1400 K for 1 ps; Then keeping half atoms in the relaxed cell fixed, we performed simulations at 2000 K (far exceeding the melting temperatures) for 100 ps, resulting in half-molten and half-crystalline models; Finally we relaxed these structures again at studied temperature for 20 ps, followed by productive simulations.   

To investigate the nucleation and growth of CdSe, larger models were used. For the brute-force MD simulation of homogeneous nucleation of CdSe, we used a supercooled liquid cell made of 8604 atoms with a box of 60.8$\times$60.3$\times$64.2 {\AA}, as illustrated in Fig.~\ref{F:initialmodel}c. This cell was obtained by first heating the crystal up to 2000 K for 200 ps and then cooling it to the studied temperature.
While for studying the growth of CdSe, a solid-liquid model consisting of 8640 atoms with a box of 39.7$\times$45.7$\times$146.3 {\AA}, as shown in Fig.~\ref{F:initialmodel}d, was used. This initial solid-liquid model was prepared in the following three steps: First, the hexagonal CdSe crystal was relaxed at 1000 K for 50 ps; Then, four layers in this cell were fixed and others were heated to 1600 K for 100 ps resulting in a solid-liquid model; Finally this model was equilibrated at studied temperature for 20 ps.

\section{Distribution of the configurational space for training and test sets}

\begin{figure}[!ht]
  \centering
  \includegraphics[width=0.9\columnwidth]{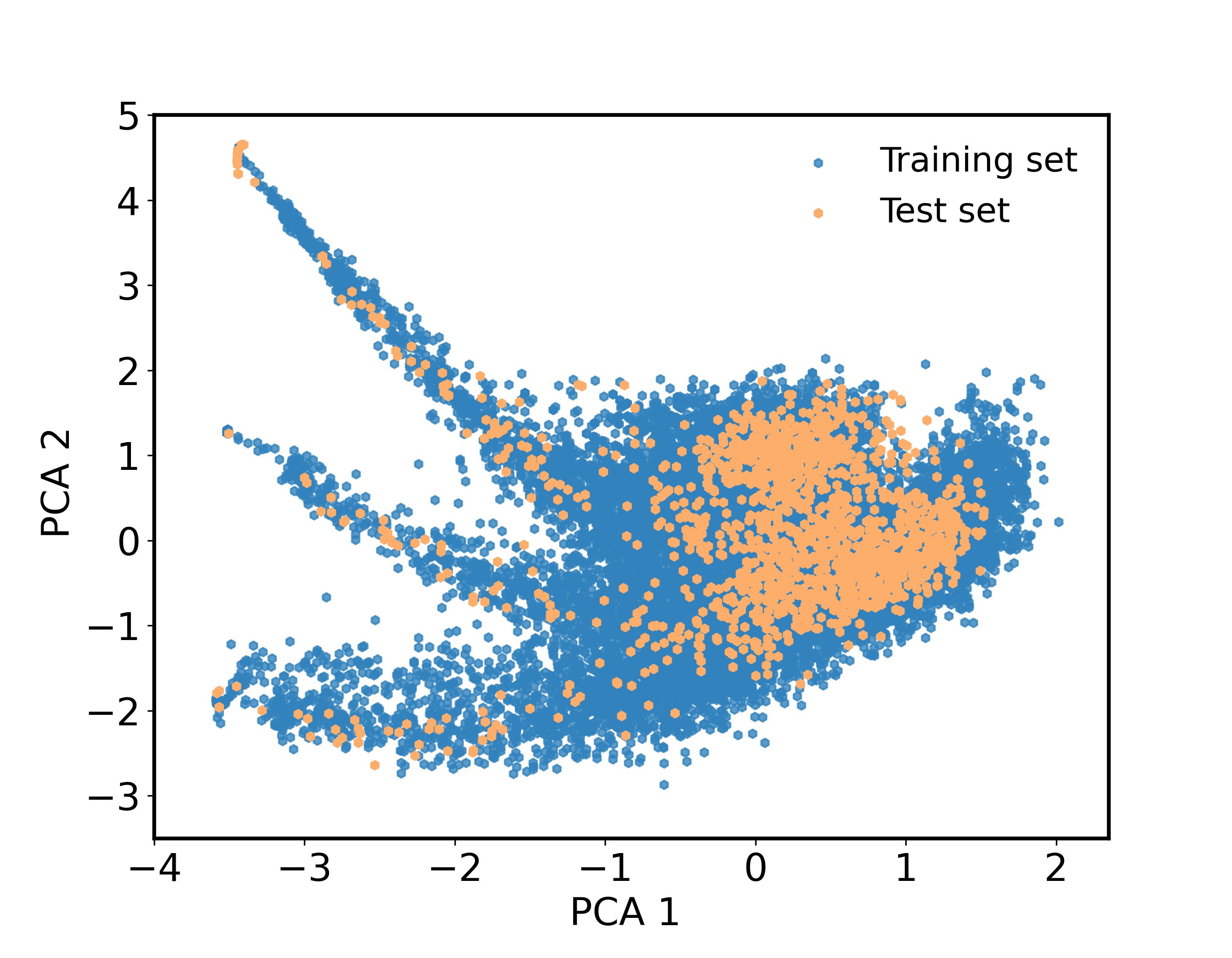}
  \caption{Comparison of the configurational space between the training and test sets. The blue and orange points represent the configurational space of training and test sets, respectively.}
  \label{F:asap}
\end{figure}

To analyze the diversity and distribution of configurations in the training and test sets, a 2D representation of the high-dimensional atomic environments of CdSe was constructed. Principal component analysis (PCA) of the local structure environment based on the smooth overlap of atomic positions (SOAP) similarity functions was performed. The test set consists of 1500 configurations  collected from DPMD and DP-WTMetaD trajectories that were performed with the final DNN potential. In these calculations, different phases of CdSe crystal (hexagonal WZ, cubic ZB, and stacking disordered phases) were heated from 300 K to 2000 K and liquid-solid coexistence structures (liquid-hexagonal WZ, liquid-cubic ZB) were heated in the temperature range of 1400 K to 2000 K. As illustrated in Fig.~\ref{F:asap}, the configurational space of the training set is much wider than that of the test set, which further confirmed the rationality and reliability of the training set. 

  \label{F:cv}
\clearpage

\section{Melting temperature}
\begin{figure}[!ht]
  \centering
  \includegraphics[width=1\columnwidth]{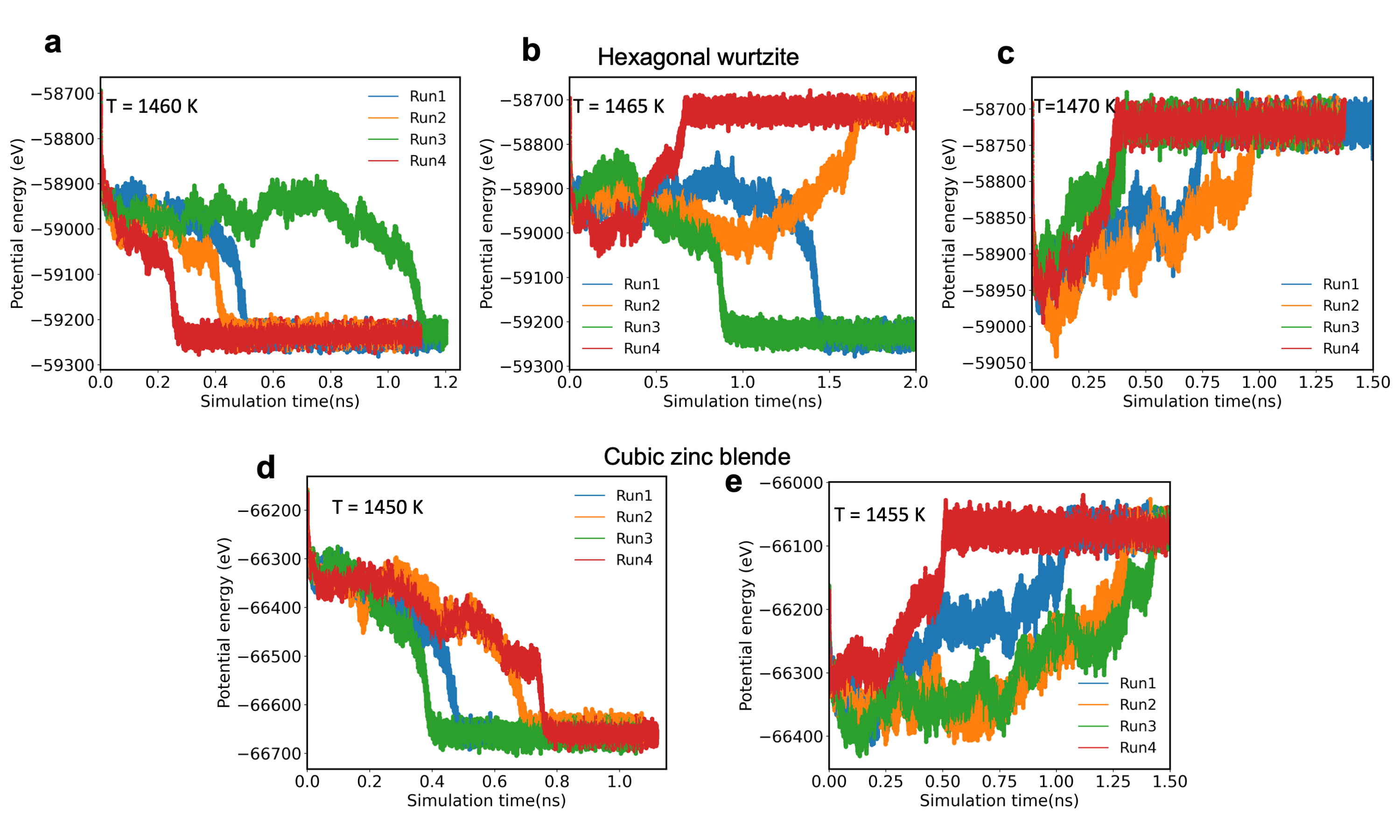}
  \caption{Evolution of the potential energy over simulation time of trajectories for studying the melting point. Wurtzite-liquid model at (a) 1460 K, (b) 1465 K, (c) 1470 K. Zinc blende-liquid at (d) 1450 K, (e) 1455 K.
   }
  \label{F:Tm}
\end{figure}

We used a two-phase coexistence method to  calculate the melting temperatures for CdSe crystals. All simulations were started with solid-liquid supercells (as described above). We estimated the melting point temperature above (or below) the simulated temperature at which all atoms in the system melt (or crystallize). The evolution of potential energy of different trajectories can be seen in Fig.~\ref{F:Tm}(a-e). The estimated melting temperatures of hexagonal and cubic phases of CdSe crystal were T$_m$ = 1465 $\pm$ 5 K and T$_m$ =  1452.5 $\pm$ 2.5 K, respectively. 

\clearpage
\section{Brute-force MD simulation of homogeneous nucleation of CdSe}
\begin{figure}[!ht]
  \centering
  \includegraphics[width=0.95\columnwidth]{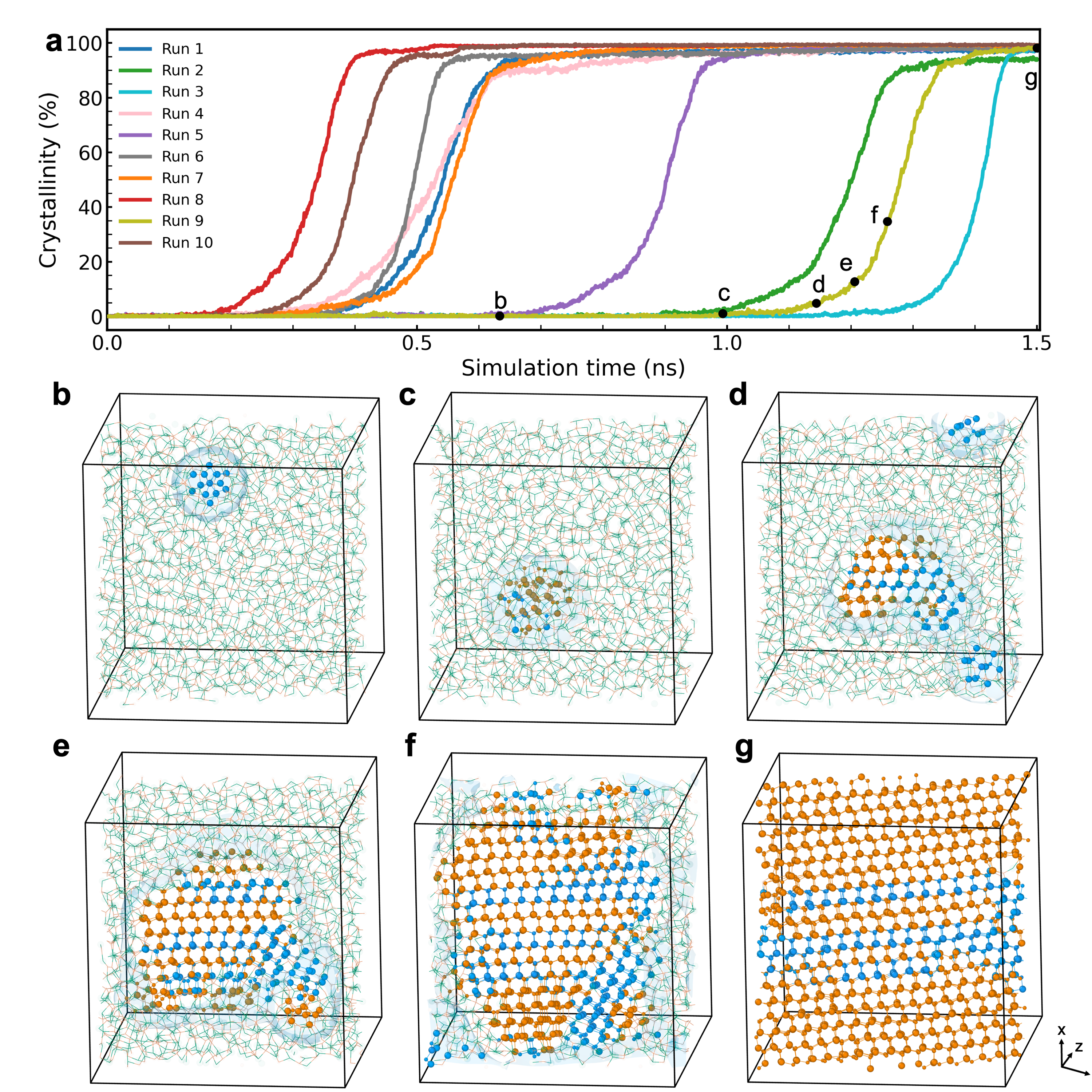}
    \caption{ Homogeneous nucleation process of CdSe at 1000 K. (a) Percentage of solid-like atoms as a function of time for 10 independent trajectories simulated at 1000 K; Typical snapshots from one of the nucleation trajectories (Run 9 in (a)) at (b) 0.63 ns, (c) 0.99 ns, (d) 1.14 ns, (e) 1.20 ns, (f) 1.25 ns and (g) 1.50 ns, respectively. Orange and blue balls represent the CdSe crystal in hexagonal WZ and cubic ZB phases, respectively. The liquid phase of CdSe is indicated by light-colored lines. The blue transparent mesh represents the interface between the liquid and solid-like phases. In this trajectory, we found that during the incubation period, a small nucleus may also occasionally emerge and then merge with the larger nucleus (d). 
    }   
  \label{F:1000k}
\end{figure}
 \clearpage
   
\begin{figure}[!ht]
  \centering
  \includegraphics[width=0.85\columnwidth]{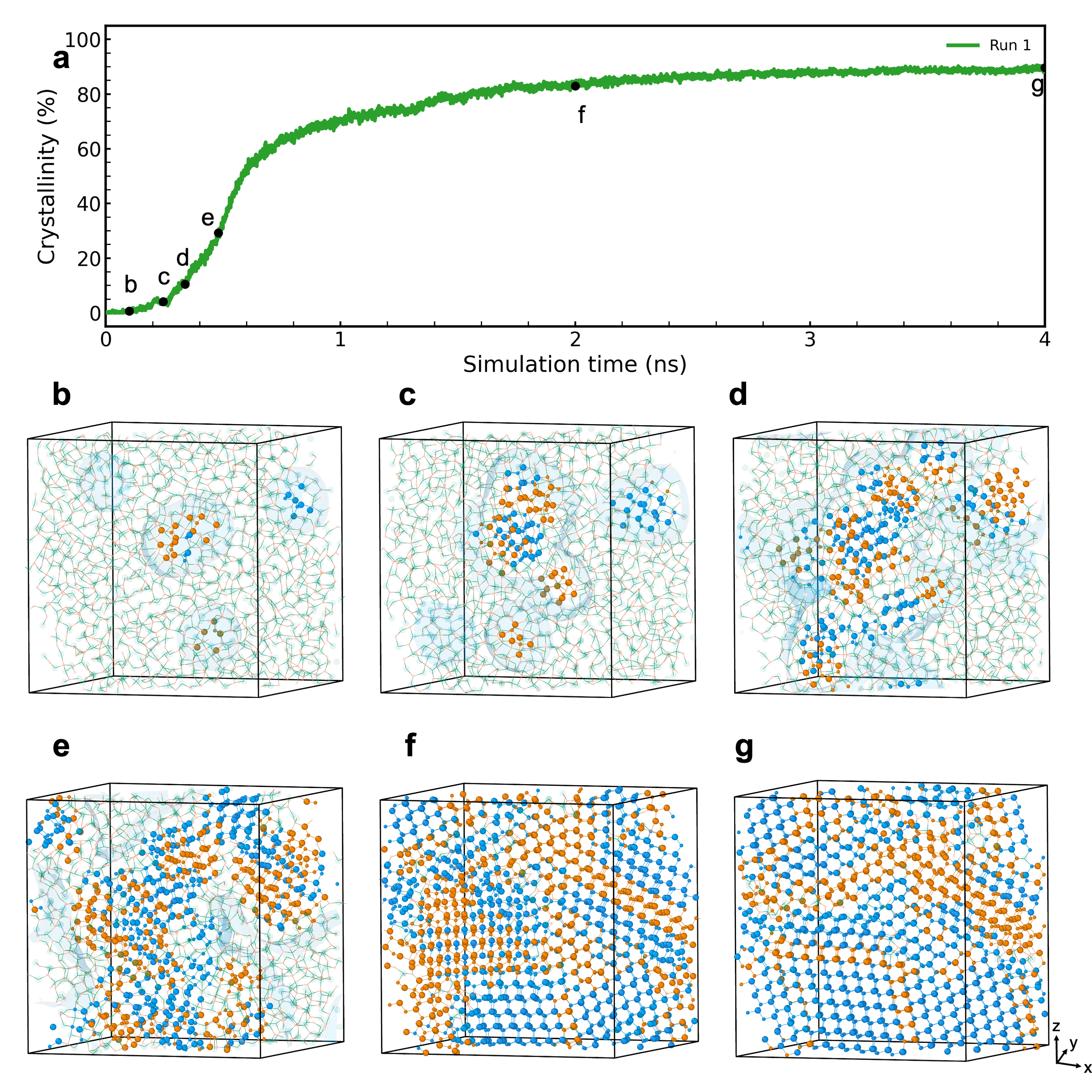}
  \caption{Homogeneous nucleation process of CdSe at 900 K. (a) Percentage of solid-like atoms as a function of time for a trajectory simulated at 900 K; Typical snapshots from the nucleation trajectories at (b) 0.1 ns, (c) 0.24 ns, (d) 0.33 ns, (e) 0.48 ns, (f) 2.0 ns and (g) 4.0 ns, respectively. Orange and blue balls represent the CdSe crystal in hexagonal WZ and cubic ZB phases, respectively. The liquid phase of CdSe is indicated by light-colored lines. The blue transparent mesh represents the interface between the liquid and solid-like phases. Contrary to the nucleation at 1000 K, we found that some small nuclei emerged and grew at 900 K during the incubation period, and formed some small crystal clusters with different orientations. The growth and interfacial adjustment of these clusters have resulted in the formation of polycrystalline structures.
  }  
  \label{F:900k}
\end{figure}
\clearpage

\begin{figure}[!ht]
  \centering
  \includegraphics[width=0.85\columnwidth]{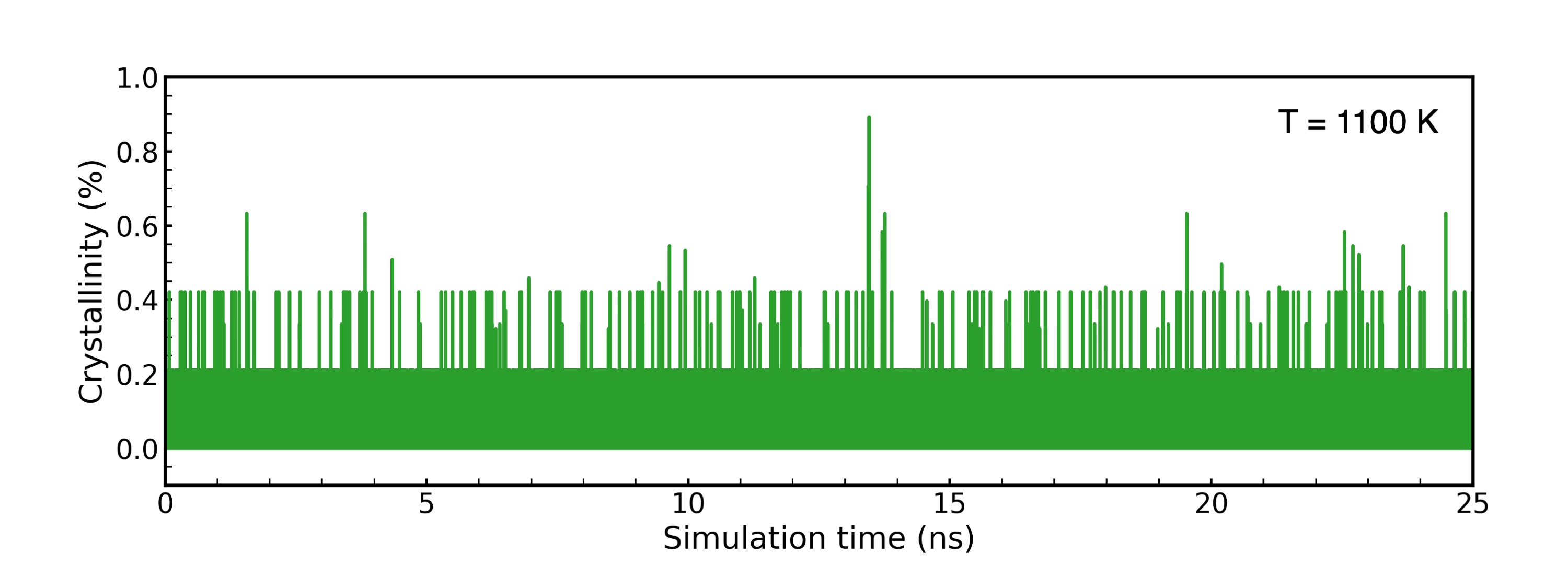}
  \caption{The percentage of solid-like atoms as a function of time for a trajectory simulated the homogeneous nucleation process of CdSe at 1100 K. No nucleation event was observed after 25 ns.
  }
  \label{F:1100}
\end{figure}
\clearpage
  
\section{Diffusion coefficients of
Se$^{2-}$ and Cd$^{2+}$}

\begin{figure}[!ht]
  \centering
  \includegraphics[width=0.85\columnwidth]{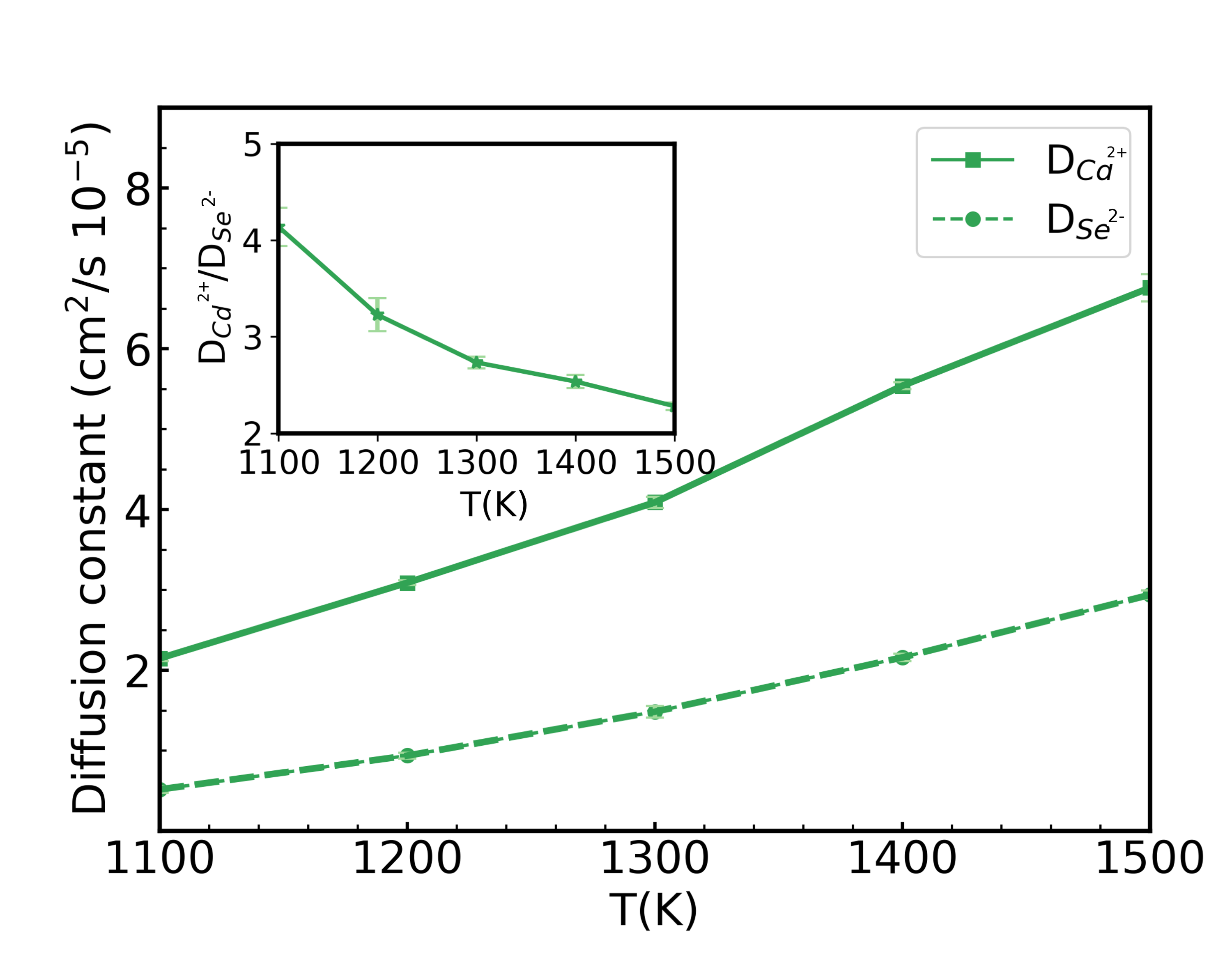}
  \caption{ Diffusion coefficients of Se$^{2-}$ and Cd$^{2+}$ as a function of temperature from 1100 K to 1500 K. The ratio of diffusion coefficients (D$_{Cd}^{2+}$/D$_{Se}^{2-}$) as a function of temperature is also given in the inset. Error bars were estimated from the standard deviation using 4 independent simulations. These simulations were run for 200 ps at each temperature.
   }
  \label{F:msd-t}
\end{figure}

\newpage

\bibliographystyle{elsarticle-num} 
\bibliography{Article}